\documentclass[a4paper,fleqn]{cas-sc}

\makeatletter
\AtEndDocument{%
  \clearpage
  \immediate\write\@auxout{\string\csxdef{lastpage}{\the\numexpr\value{page}-1\relax}}%
}
\makeatother

\usepackage[utf8]{inputenc}
\usepackage[numbers,sort&compress]{natbib}
\usepackage{amsmath}
\usepackage{amssymb}   % \lesssim, \gtrsim -- do not drop
\usepackage{booktabs}
\usepackage{graphicx}
\graphicspath{{figures/}}
\usepackage{xcolor}
\usepackage{xurl}
\usepackage{xr}
\usepackage{listings}
\usepackage{enumitem} 

\lstdefinestyle{samqeq}{
  basicstyle=\ttfamily\footnotesize,
  commentstyle=\itshape\color{black!60},
  keywordstyle=\bfseries,
  showstringspaces=false,
  columns=fullflexible,
  keepspaces=true,
  breaklines=true,
  breakatwhitespace=true,
  breakindent=2em,
  aboveskip=0.8\baselineskip,
  belowskip=0.6\baselineskip,
  xleftmargin=1em,
}
\lstdefinelanguage{LAMMPS}{
  morekeywords={fix,unfix,fix_modify,compute,pair_style,pair_coeff,kspace_style,
                units,atom_style,group,set,thermo_style,run,timestep,velocity},
  morecomment=[l]{\#},
  sensitive=true,
}

\makeatletter
\newcommand*{\myexternaldocument}[2][]{%
  \externaldocument[#1]{#2}%
}
\makeatother
\myexternaldocument[SI-]{si-xref}

\newcommand{\qvec}{\mathbf{q}}
\newcommand{\Jij}{J_{ij}}
\newcommand{\Amat}{\widetilde{\mathbf{A}}}
\newcommand{\Pmat}{\mathbf{P}}
\newcommand{\Hmat}{\mathbf{H}}

\newcommand{\lmin}{\lambda_{\mathrm{min}}}
\newcommand{\qzeroi}[1]{q^{0}_{#1}}
\newcommand{\code}[1]{\texttt{#1}}         % C++ identifiers, calls, file names
\newcommand{\kw}[1]{\texttt{#1}}           % LAMMPS input-script keywords/values
\newcommand{\samQEq}{\textit{samQEq}}      % this program
\newcommand{\ncases}{22}

\begin{document}

\let\WriteBookmarks\relax
\def\floatpagepagefraction{1}
\def\textpagefraction{.001}

\shorttitle{samQEq for molecules and conductors}
\shortauthors{Solan, Sun and Pascal}

\title[mode=title]{samQEq: universal charge equilibration for molecules and conductors}

\author[1]{Nicholas Solan}[orcid=0009-0007-8884-4249]
\fnmark[1]

\author[1]{David Sun}[orcid=0009-0009-8977-0151]
\fnmark[1]

\author[1,2]{Tod A. Pascal}[orcid=0000-0003-2096-1143]
\cormark[1]
\ead{tpascal@ucsd.edu}
\affiliation[1]{organization={ATLAS Materials Physics Laboratory, Aiiso Yufeng Li
            Family Department of Chemical and Nano Engineering,
            University of California San Diego},
            city={La Jolla},
            citysep={,},
            postcode={92093},
            state={CA},
            country={USA}}
\affiliation[2]{organization={Program in Materials Science and Engineering,
            University of California San Diego},
            city={La Jolla},
            citysep={,},
            postcode={92093},
            state={CA},
            country={USA}}
\cortext[1]{Corresponding author}
\fntext[1]{These authors contributed equally to this work.}

\begin{abstract}
Charge equilibration assigns atomic charges by equalizing an atomic chemical potential built from electronegativity, hardness and the Coulomb interaction. 
Under one global constraint it lets charge flow without limit between molecules, which fluctuating-charge 
models prevent with one neutrality constraint per molecule. A metal needs the opposite: a global constraint 
and a hardness near zero, and there the charge operator becomes indefinite unless it includes each charge's own self-energy.
Here, we present \samQEq{} (solids-and-molecules QEq), a unified charge-equilibration package for LAMMPS that 
combines the two constructions in one energy functional and one matrix-free solve. An orthogonal projector 
imposes one neutrality constraint per molecule, and the finite-width self-energy of a Gaussian charge on 
designated conductor types, narrower than the conductor's own Gaussian, keeps the conductor's part of the operator positive definite at any hardness. On a gold slab carrying a water 
film, the operator stays definite down to the conductor limit of the model, the metal screens 
a point charge by a constant amount at every height beyond 4~\AA{}, and the film's charges stay bounded under dynamics. Reciprocal-space electrostatics is evaluated inside the 
matrix-vector product, and three distinct shielding kernels are run-time settings. 
We verify the implementation by reproducing a published fluctuating-charge water model without refitting, 
and the native ReaxFF charge-equilibration step to approximately $10^{-9}\,e$. On a single four-V100 node the
largest tested systems run at approximately twice the throughput of a 128-core CPU node. \samQEq{} thus 
provides one charge-equilibration layer for molecules, conductors and the interfaces between them.
\end{abstract}

\begin{highlights}
\item Charge equilibration that is well posed for both molecules and conductors
\item A narrow Gaussian self-energy keeps the conductor block definite at any hardness
\item One neutrality constraint per molecule, or one global constraint, in one solver
\item Particle-mesh Ewald inside the matrix-vector product, with three shielding kernels
\item Kokkos device kernels for the charge solve on a single GPU node
\end{highlights}


\maketitle

\vspace{1em}
\noindent\rule{\linewidth}{0.5pt}
\begin{description}
\item[Program title:] samQEq
\item[CPC Library link to program files:] (to be added by Technical Editor)
\item[Developer's repository link:] \url{https://github.com/atlas-nano/codes/tree/main/samqeq}
\item[Archived version:] samQEq 1.0.2, \url{https://doi.org/10.5281/zenodo.23629400}
\item[Code Ocean capsule:] (to be added by Technical Editor)
\item[Licensing provisions:] GPL-2.0 (derived from LAMMPS)
\item[Programming language:] C++17, with optional Kokkos device kernels
\item[Supplementary material:] The per-fragment projector, the solvers and their convergence
  criteria, the dense diagonalization of the constrained operator, the parameter sets, the simulation
  protocols, the regression suite, the device benchmark and the
  charge-solve tolerance study.
\item[Nature of problem:] Atomic charges that respond to their environment are obtained by minimizing a
  second-order charge-equilibration functional [1,2] under neutrality constraints, in periodic systems with
  long-range electrostatics, at every step of a molecular dynamics run. Under one global constraint the
  functional lets charge flow without limit between atoms, appropriate for conductors but not molecules.[3] 
  At the near-zero hardness of a conductor [4] the operator of a kernel without the charge's
  self-energy becomes indefinite.[5] The computational problem is a single linear solve in which
  molecules and conductors are both well posed, with forces consistent with the solved charges.
\item[Solution method:] We implement a single second-order charge-equilibration energy functional as a
  LAMMPS fix and make both limits well posed within one solve. An orthogonal projector imposes one
  neutrality constraint per molecule [6] (the insulating limit), or a single global constraint, and for
  designated conductor types the finite-width self-energy of a Gaussian charge is added to the on-site
  hardness [4], which, for a self-energy narrower than the type's own Gaussian, keeps the conductor's part of the
  operator positive definite at any hardness. The reciprocal-space contribution is evaluated inside the matrix-vector product [5,7,8], 
  so that one projected Krylov solve returns fully long-range charges with no self-consistent-field outer loop. 
  The cube-root, Gaussian-overlap [9] and exact Slater-orbital shielding kernels are run-time settings, 
  mirrored by a companion pair style so that the forces and the charge solve use one kernel. The fix and 
  its pair style are overlaid on any other LAMMPS potential. Charges may be minimized at each step, 
  propagated by an extended Lagrangian, or advanced by a time-reversible predictor--corrector.
\item[Additional comments including Restrictions and Unusual features:]
  \emph{External routines.} The package builds within a LAMMPS source tree [10], tested at version
  2~Sep~2026, with the long-range electrostatics, molecule, rigid-body and extra-fix packages enabled. 
  The optional device kernels additionally require the Kokkos package. Three edits to the host tree are needed, 
  and all three ship with the package as patches: the registration of the package in the build system's
  package list; a zeroing of the per-atom particle-mesh accumulators, without which a per-atom energy can
  be read from uninitialized memory; and a friend declaration in the external-field fix header, so that the
  base class can read the applied field. The distributed build recipe and regression suite are included. 
  The solver itself is self-contained: the augmented saddle-point solver and the per-atom reciprocal 
  particle-particle particle-mesh routine are both shipped in the package, so the package does not depend on the ReaxFF package.
  \emph{Restrictions.} Conductors carry a fixed total charge; there is no constant-potential mode. 
  The bond-softness kernel restores linear scaling of the polarizability with chain length but needs
  additional parameters. MINRES is unpreconditioned. The extended-Lagrangian and predictor--corrector 
  charge propagators do not conserve energy at the documented settings. The device solve 
  only covers the atomic-charge path on a single node.
  \emph{Features.} Correctness is enforced by \ncases{} self-contained regression cases compared against 
  committed per-atom charges at $10^{-9}$ relative tolerance, run from an automated pre-commit test hook. 
  The combined molecule--conductor test is covered by two of the regression cases,
  each with its decks and reference output. Section~\ref{sec:checks} 
  documents the initialization checks that detect when the solved operator differs from the one the input specifies.
\item[References:]
  \begin{enumerate}
  \def\labelenumi{[\arabic{enumi}]}
  \item W.~J.~Mortier, S.~K.~Ghosh, S.~Shankar, Electronegativity-equalization method for the calculation of atomic charges in molecules, J. Am. Chem. Soc. 108 (1986) 4315--4320. \url{https://doi.org/10.1021/ja00275a013}
  \item A.~K.~Rapp\'e, W.~A.~Goddard~III, Charge equilibration for molecular dynamics simulations, J. Phys. Chem. 95 (1991) 3358--3363. \url{https://doi.org/10.1021/j100161a070}
  \item T.~Verstraelen, P.~W.~Ayers, V.~Van~Speybroeck, M.~Waroquier, ACKS2: Atom-condensed Kohn--Sham DFT approximated to second order, J. Chem. Phys. 138 (2013) 074108. \url{https://doi.org/10.1063/1.4791569}
  \item L.~Scalfi, T.~Dufils, K.~G.~Reeves, B.~Rotenberg, M.~Salanne, A semiclassical Thomas--Fermi model to tune the metallicity of electrodes in molecular simulations, J. Chem. Phys. 153 (2020) 174704. \url{https://doi.org/10.1063/5.0028232}
  \item K.~Savvidi, L.~J.~V.~Ahrens-Iwers, L.~Colombi~Ciacchi, D.~Zahn, M.~H.~M{\"u}ser, R.~H.~Mei{\ss}ner, Efficient treatment of long-range electrostatics in charge equilibration approaches, J. Chem. Phys. 162 (2025) 174108. \url{https://doi.org/10.1063/5.0255823}
  \item S.~W.~Rick, S.~J.~Stuart, B.~J.~Berne, Dynamical fluctuating charge force fields: Application to liquid water, J. Chem. Phys. 101 (1994) 6141--6156. \url{https://doi.org/10.1063/1.468398}
  \item L.~J.~V.~Ahrens-Iwers, R.~H.~Mei{\ss}ner, Constant potential simulations on a mesh, J. Chem. Phys. 155 (2021) 104104. \url{https://doi.org/10.1063/5.0063381}
  \item M.~Gubler, J.~A.~Finkler, M.~R.~Sch{\"a}fer, J.~Behler, S.~Goedecker, Accelerating fourth-generation machine learning potentials using quasi-linear scaling particle mesh charge equilibration, J. Chem. Theory Comput. 20 (2024) 7264--7271. \url{https://doi.org/10.1021/acs.jctc.4c00334}
  \item S.~Naserifar, D.~J.~Brooks, W.~A.~Goddard~III, V.~Cvicek, Polarizable charge equilibration model for predicting accurate electrostatic interactions in molecules and solids, J. Chem. Phys. 146 (2017) 124117. \url{https://doi.org/10.1063/1.4978891}
  \item A.~P.~Thompson, H.~M.~Aktulga, R.~Berger, D.~S.~Bolintineanu, W.~M.~Brown, P.~S.~Crozier, P.~J.~in~'t~Veld, A.~Kohlmeyer, S.~G.~Moore, T.~D.~Nguyen, R.~Shan, M.~J.~Stevens, J.~Tranchida, C.~Trott, S.~J.~Plimpton, LAMMPS -- a flexible simulation tool for particle-based materials modeling at the atomic, meso, and continuum scales, Comput. Phys. Commun. 271 (2022) 108171. \url{https://doi.org/10.1016/j.cpc.2021.108171}
  \end{enumerate}
\end{description}
\noindent\rule{\linewidth}{0.5pt}
\vspace{1em}

% =====================================================================
\section{Introduction}
% =====================================================================

The electrostatic energy in molecular simulations depends on the charge distribution, and that distribution 
is not fixed: a molecule polarizes in the field of its neighbors, an ion redistributes charge with its 
solvation shell, and a metal surface screens an applied field. Fixed-charge models cannot represent this 
environmental response, whereas charge equilibration allows the atomic charges to vary with configuration. 
In the electronegativity equalization method (EEM) of Mortier, Ghosh and Shankar\cite{Mortier1986} and the charge
equilibration (QEq) scheme of Rapp\'e and Goddard,\cite{Rappe1991} the atomic charges are obtained by
equalizing an atomic chemical potential constructed from experimental ionization potentials, electron
affinities and shielded electrostatic interactions. The charges then depend on geometry and respond to
the environment, at the cost of repeated electrostatic evaluations within each step. Patel and Brooks
carried the approach into a biomolecular force field,\cite{Patel2004} while Naserifar and co-workers
replaced the point charge by a Gaussian of element-specific width whose shielded interaction stays 
finite at every separation.\cite{Naserifar2017}

The limitations of the original construction are equally well characterized. Because the chemical 
potential is equalized over the entire system, charge flows without limit, and 
every system behaves as a metal with respect to charge transfer. Chen and Mart\'inez showed that this yields incorrect 
asymptotics when bonds are broken and introduced QTPIE, which is formulated in charge-transfer
variables and attenuates the transfer with orbital overlap.\cite{Chen2007} Split-charge equilibration 
reaches the same end through a distance-dependent bond hardness.\cite{Nistor2006} Verstraelen and 
co-workers derived ACKS2 from Kohn--Sham density functional theory, restoring the linear size-dependence 
of the polarizability and the correct dissociation limit through constrained atomic populations and 
a Legendre-transformed kinetic energy.\cite{Verstraelen2013} Rick and co-workers applied an alternative approach 
to liquid water: a per-molecule neutrality constraint.\cite{Rick1994} More 
recently, Gergs and co-workers generalized QEq and QTPIE into a model that constrains the 
split-charge transfer locally without specifying a bond hardness.\cite{Gergs2021}

At the other end, global electronegativity equalization is metallic by construction,\cite{Nistor2009} and metals 
need a hardness near zero: under a global constraint 
the charge-equilibration functional is the semiclassical Thomas--Fermi model, with the electronegativity 
in the role of the Fermi level and the hardness in that of the screening length.\cite{Scalfi2020} 
Yet, as the QEq hardness approaches zero, the operator loses positive definiteness, and the solve
returns a saddle point or fails to converge.\cite{Savvidi2025} Various remedies have been proposed: 
employing constant-potential electrode models, which represent each electrode atom by a Gaussian charge whose self-energy
sits on the diagonal,\cite{Siepmann1995,Reed2007} adding a Thomas--Fermi hardness term,\cite{Scalfi2020}
or rederiving the polarizable electrode from chemical-potential equalization itself.\cite{Nakano2019} 
Gaussian charges with the self-term on the hardness have been employed in QEq force fields
for metal oxides,\cite{Streitz1994} recent machine-learned potentials,\cite{Ghasemi2015,Ko2021} 
and to electrochemical cells under applied potentials.\cite{Onofrio2015} 

Recently introduced machine-learned interatomic potentials (MLIPs) face a related limitation in treating their electrostatics. 
Most are short-ranged by construction and require message passing frameworks to extend their reach 
for condensed-phase simulations. Alternatively, fourth-generation neural-network potentials employ QEq-like schemes, 
predicting atomic electronegativities 
and solving under a global charge constraint with cutoffs\cite{Ko2021} and more recently on a particle mesh.\cite{Gubler2024}
Another approach is the Latent Ewald Summation formalism, where a hidden per-atom variable is learned from 
the local descriptors and summed in reciprocal space with a Gaussian smearing factor.\cite{Cheng2025,Kim2025LES} 
The field response of such a model follows in the linear regime from the Born effective charges.\cite{Zhong2025} 
Fragment-resolved constraints on a learned charge equilibration model have also been introduced for electrochemical 
interfaces, with the fragments themselves a learned, differentiable function of the geometry, trained on DFT charges 
for an oxide electrode in water.\cite{Peeketi2026} Notably, in each case, the electrostatics is part of what the model encodes. 

For practical MD simulations, QEq options are available as standalone packages in LAMMPS with point, shielded, Slater and dynamic shielding variants, 
as the charge steps of COMB, ReaxFF, ACKS2 and QTPIE, and as the constant-potential ELECTRODE package.\cite{Thompson2022,AhrensIwers2022,phillpot2018} 
Each prescribes the kernel, the long-range treatment and the constraint together (Section~\ref{sec:comparison}). 
The ELECTRODE package solves the electrode charges on the particle mesh under the field of a fixed-charge electrolyte;\cite{AhrensIwers2022} 
MetalWalls, outside LAMMPS, solves them together with the induced dipoles of a polarizable electrolyte.\cite{Coretti2022} 
None solves a fluctuating-charge electrolyte, one neutrality constraint per molecule, in the same charge-equilibration solve as the conductor. 

What is missing is an abstract electrostatics layer that any potential can call, with a physical energy, a choice of kernel, 
full long-range sums, and constraints that span molecules and metals. To that end, we developed \samQEq{}, or solids-and-molecules QEq, 
a universal charge-equilibration package for LAMMPS\cite{Thompson2022} that combines the two constructions in one energy functional 
and one solve (Figure~\ref{fig:arch}). An orthogonal projector imposes one total-charge constraint per molecule, which enforces the
insulating limit: charge redistributes within each molecule but not between molecules. A finite-width
self-energy of a Gaussian charge enforces the conductor limit: for a width narrower than the conductor's own Gaussian it keeps
the conductor's part of the operator positive definite at any hardness. One projected Krylov solve satisfies both, with the reciprocal-space term inside the
matrix-vector product. The shielding kernel and the long-range treatment are run-time settings; every kernel is used as defined, and
the particle-mesh sum inside the matrix-vector product is the same for all of them. This allows simultaneous treatment of 
molecules, metals and \emph{molecules on metals}. On a gold slab carrying a water film we show that, of the four combinations of constraint and self-energy, only this 
one is definite at every hardness without charge transfer between molecules (Section~\ref{sec:twolimits}). 

The remainder of the paper is organized as follows: Section~\ref{sec:theory} gives the functional, the kernels, the constrained 
solve and the conductor limit. Section~\ref{sec:longrange} describes the evaluation of the reciprocal-space term inside the 
matrix-vector product, and Section~\ref{sec:coupling} the composition with other potentials. Section~\ref{sec:implementation} 
describes the code architecture and the run-time controls, while Section~\ref{sec:stability} goes beyond implementation 
details to discuss the conditions under which the solve is well posed and the guards that keep it so. Section~\ref{sec:validation} 
reports the validation: molecules and a conductor in one solve, the central result, then the reproduction of a 
published water model and of the ReaxFF charge step with no parameter refit. Section~\ref{sec:limitations} states the current limitations, 
Section~\ref{sec:performance} reports performance, and Section~\ref{sec:comparison} positions the method against related approaches.

% =====================================================================
\section{Theory}
\label{sec:theory}
% =====================================================================

\subsection{The energy functional}

We reference each charge to a per-type formal charge,
\begin{equation}
q_i \;=\; \qzeroi{i} + \Delta_i ,
\label{eq:qref}
\end{equation}
where $\qzeroi{i}$ is the reference charge of the type of atom $i$ and
$\Delta_i$ the solved deviation. Following the standard expansion of the energy in the atomic charges,\cite{Mortier1986,Rappe1991} which 
is first order in the electronegativity $\chi_i$ and second order in the hardness $\eta_i$, we write
\begin{equation}
E(\qvec) \;=\; \sum_i \left[ \chi_i q_i + \tfrac{1}{2}\eta_i \left(q_i - \qzeroi{i}\right)^2 \right]
\;+\; \tfrac{1}{2} \sum_{i \neq j} \Jij(r_{ij})\, q_i q_j ,
\label{eq:functional}
\end{equation}
where $q_i$ is the charge on atom $i$, $\chi_i$ is taken as
$\tfrac{1}{2}(\mathrm{IP}+\mathrm{EA})$, $\eta_i$ is taken as
$\mathrm{IP}-\mathrm{EA}$, with IP and EA the ionization potential and electron affinity of the atom, and
$\Jij(r_{ij})$ is a shielded Coulomb kernel between atoms $i$ and $j$ separated by $r_{ij}$. The reference 
places the center of the on-site parabola at the specified formal charge rather than at zero: an isolated atom takes the charge $\qzeroi{i} - \chi_i/\eta_i$. 
The projector of Section~\ref{sec:constraint} holds each fragment's total charge at the sum of its reference charges, which we refer to throughout as the
``neutrality'' constraint. For $\qzeroi{i} = 0$, the case of every model in Section~\ref{sec:validation}, Eq.~\eqref{eq:functional} is the familiar 
unreferenced functional.

\subsection{Shielding kernels}
\label{sec:kernels}

The off-diagonal kernel is the point at which published models differ most, and we therefore make it selectable at run time rather than fixed at compile time
(Table~\ref{tab:kernels}). All three tabulated forms approach $1/r$ at large separation, so the Ewald complement is unchanged, and all three are
dimensionally consistent in both supported unit systems without a conversion factor.

The exact Slater form allows published models to be used without a refit. Published Slater-overlap fluctuating-charge models are parameterized against the exact
Slater-orbital Coulomb integral,\cite{Rick1994} so pinning such a model to an approximate kernel requires a refit, and a refit obscures whether an
implementation reproduces the original. With the exact integral available, the original exponents may be used directly (Section~\ref{sec:port}).

\begin{table}[t]
\caption{Interchangeable off-diagonal shielding kernels. The fourth column of the parameter file is reinterpreted according to the selected form. The
companion pair style accepts the same selection, so that the forces and the charge solve use one kernel. The Slater form is the closed-form two-center 
Coulomb integral over $1s$ or $2s$ Slater densities given by Rick and co-workers;\cite{Rick1994} we evaluate it by tabulation, verified against direct
numerical quadrature in Section~\ref{sec:port}. The Gaussian form is that of Naserifar and co-workers,\cite{Naserifar2017} with the pair exponent of Eq.~\eqref{eq:betaij}.}
\label{tab:kernels}
\begin{tabular}{lll}
\toprule
Kernel & $\Jij(r)$ & Parameter \\
\midrule
\code{cbrt} (default) & $\left(r^3 + \gamma^{-3}\right)^{-1/3}$ & $\gamma$ (inverse length) \\
\code{gaussian}       & $\mathrm{erf}(\beta_{ij} r)/r$, $\alpha_i = \lambda/(2R_{c,i}^2)$ & $R_c$ (Gaussian radius) \\
\code{slater}         & exact Slater-orbital integral\cite{Rick1994} & $\zeta$ (Slater exponent) \\
\bottomrule
\end{tabular}
\end{table}

The conductor mode and the machine-learning correspondence of this paper rest on the Gaussian form. It is the Coulomb interaction of two normalized Gaussian charge densities,
$\rho_i(\mathbf{r}) \propto \exp(-\alpha_i r^2)$, with the per-type exponent $\alpha_i = \lambda/(2R_{c,i}^2)$,\cite{Naserifar2017} where $R_{c,i}$ is the 
Gaussian radius of the type and $\lambda$ a dimensionless width factor. The kernel depends on the pair only through
\begin{equation}
\beta_{ij} \;=\; \left( \frac{\alpha_i \alpha_j}{\alpha_i + \alpha_j} \right)^{1/2} ,
\label{eq:betaij}
\end{equation}
where $\beta_{ij}$ is an inverse length, since $\alpha_i$ has units of (length)$^{-2}$. In \samQEq{}, the combination rule may be overridden for any type pair 
at run time. Because the kernel is the interaction of a real charge distribution, its reciprocal-space form is exact rather than fitted,
\begin{equation}
\int \mathrm{d}^3 r \; e^{-i\mathbf{k}\cdot\mathbf{r}}\,
\frac{\mathrm{erf}(\beta_{ij} r)}{r}
\;=\; \frac{4\pi}{k^2}\, \exp\!\left( -\frac{k^2}{4\beta_{ij}^2} \right) ,
\label{eq:gauss-fourier}
\end{equation}
where $\mathbf{k}$ is a wavevector of magnitude $k$. The shielding is therefore equivalent to a Gaussian factor on each wavevector of the Ewald sum. For a single width, 
the factor is $\exp(-\sigma^2 k^2/2)$ with $\sigma = 1/(\sqrt{2}\,\beta_{ij})$, which is the smearing factor of Latent Ewald Summation;\cite{Cheng2025} 
Section~\ref{sec:mlip} returns to that correspondence.
In the cube-root form the pair parameter is $\gamma_{ij} = (\gamma_i \gamma_j)^{1/2}$; in the Slater form each type carries either the $1s$ or the $2s$ 
Slater form factor. 

For designated ion pairs the off-diagonal element may additionally be Tang--Toennies damped on top of the active kernel,\cite{Tang1984}
\begin{equation}
\Jij^{\rm damp}(r) = f_n(b\,r)\,\Jij(r),
\qquad
f_n(x) = 1 - e^{-x}\sum_{k \le n} \frac{x^k}{k!},
\label{eq:tt}
\end{equation}
where $f_n$ is the Tang--Toennies damping function of order $n$ (default 4) and $b$ is the damping range parameter in \AA$^{-1}$. 
Since $f_n \to 0$ at contact,
the damping suppresses the contact charge-transfer coupling, which otherwise transfers too much charge across a contact ion pair.
Additionally, since $f_n \to 1$ at large $r$, it leaves the long-range complement unchanged. 
Every kernel choice is mirrored in the companion pair style.

\subsection{Enforcing the molecular limit}
\label{sec:constraint}

We minimize the charges using Lagrange multipliers (i.e, subject to a constraint applied per fragment) over the atoms of a designated group $\mathcal{G}$. 
Writing $\Hmat$ for the Hessian of Eq.~\eqref{eq:functional} and $\Pmat$ for the orthogonal projector onto the constrained subspace, the solve is
\begin{equation}
\Amat\, \boldsymbol{\Delta} \;=\; \Pmat\,\mathbf{b},
\qquad
\Amat \;=\; \Pmat\,\Hmat\,\Pmat ,
\qquad
b_i = -\chi_i - \sum_{j \neq i} \Jij\, \qzeroi{j} ,
\label{eq:solve}
\end{equation}
with $\Hmat$ restricted to $\mathcal{G}\times\mathcal{G}$, so that the reference charges enter as a fixed field. For atoms $j$ outside $\mathcal{G}$, the sum in
$b_i$ uses their actual fixed charges $q_j$ in place of $\qzeroi{j}$. For per-fragment
neutrality the projector removes each fragment's uniform-charge direction,
\begin{equation}
(\Pmat \mathbf{v})_i \;=\; v_i \;-\; \frac{1}{N_{\mathcal{M}(i)}}
\sum_{j \in \mathcal{M}(i)} v_j ,
\label{eq:projector}
\end{equation}
where $\mathcal{M}(i)$ is the fragment containing atom $i$, $N_m$ the number of atoms in fragment $m$, and $\mathbf{v}$ any per-atom vector. 
Rather than eliminating the constraint, we project each search direction onto the constrained subspace, which keeps the operator symmetric and makes
the constraint a property of the projector rather than of the linear system. Alternative constraint forms are therefore run-time settings that reuse the
same solver. Here we use per-fragment neutrality or global neutrality throughout.

Per-fragment neutrality is the constraint of the fluctuating-charge model of Rick and co-workers:\cite{Rick1994} charge redistributes within a molecule but
not between molecules, which is the correct limit for an insulating molecular liquid. In the absence of fragment identifiers, the projector degenerates to
global charge neutrality, recovering the standard electronegativity-equalization constraint appropriate to a conductor (Section~\ref{sec:conductor}). 
Both limits are reached without a change of code path.

\subsection{The conductor limit}
\label{sec:conductor}

Under the global constraint, Eq.~\eqref{eq:functional} is equivalent to the semiclassical Thomas--Fermi electrode model of Scalfi and coworkers, who noted that the two 
self-consistent problems coincide when $\chi$ plays the role of the Fermi level and $\eta$ that of $e^2 l_{\rm TF}^2 d/\varepsilon_0$, with $l_{\rm TF}$ the
Thomas--Fermi screening length, $d$ the atomic density and $\varepsilon_0$ the vacuum permittivity.\cite{Scalfi2020} Whether the limit is well posed 
depends on how the finite size of the charge enters the solve. For a molecule, the fitted $\eta$ already contains the electrostatic cost of confining charge to an 
atom of finite size, and the operator diagonal therefore carries no self-energy; the grid self-term of Section~\ref{sec:longrange} is subtracted to keep it so. 
For a metal, by contrast, $\eta$ is a bulk quantity with no width, and the finite size must be supplied separately. We supply it, for designated types, 
as the diagonal element of the Gaussian kernel,
\begin{equation}
\eta_i \;\rightarrow\; \eta_i + K_e \left( \frac{2\alpha_w}{\pi} \right)^{1/2},
\qquad \alpha_w = \frac{1}{2w^2} ,
\label{eq:gself}
\end{equation}
where $w$ is the standard deviation of the Gaussian charge and $K_e$ the Coulomb constant. The added term is the $r \to 0$ limit of $\mathrm{erf}(\beta_{ii} r)/r$
for a Gaussian of width $w$. For the Gaussian shielding kernel this gives a rigorous sufficient condition for the solve to be well posed. Let $J_{ii}(0) = K_e (2\alpha_i/\pi)^{1/2}$ be
the self-energy of the type's own Gaussian. The kernel matrix completed by $J_{ii}(0)$ on its diagonal is the interaction matrix of a set of real charge densities, 
a Gram matrix, and is positive semidefinite, so
\begin{equation}
\lambda_{\min}\!\left(\Pmat\Hmat\Pmat\,\big|_{\mathrm{range}\,\Pmat}\right) \;\ge\; \min_i \left( \eta_i + s_i - J_{ii}(0) \right),
\label{eq:gram}
\end{equation}
where $s_i$ is the self-energy of Eq.~\eqref{eq:gself} on the conductor types and zero on the others. The bound holds for the full operator 
and therefore for its restriction to the constrained subspace, since the operator $\Pmat\Hmat\Pmat$ itself has one exact zero 
eigenvalue per fragment, on the null space of the projector. The argument requires every off-diagonal element to be the 
Coulomb interaction of the two sites' own Gaussian densities: the Gaussian kernel with its combination rule, Eq.~\eqref{eq:betaij}, 
shielding every pair. A per-pair radius override, a damped ion pair, or bare intermolecular pairs 
(the intramolecular-shielding mode of Section~\ref{sec:longrange}) leave the Gram form, and the bound no longer applies. 

The bound is also a minimum over all sites. A molecular site whose hardness is below its own contact value makes it negative 
and uninformative. Thus, for an interface Eq.~\eqref{eq:gram} removes the instability that the conductor would introduce, 
while the definiteness of the combined operator is established by dense diagonalization (Section~\ref{sec:twolimits}). 
Without the self-energy, a conductor's operator can become indefinite once its hardness falls below the contact value of its own kernel. 
With it, the smallest eigenvalue is bounded below by $\eta_i + K_e[(\sqrt{\pi}w)^{-1} - (2\alpha_i/\pi)^{1/2}]$, 
positive for every $\eta_i \ge 0$ whenever $w$ is narrower than the type's own Gaussian, 
$(2\alpha_i)^{-1/2}$. At equal widths the bound reaches zero as $\eta \to 0$, where the Gram matrix of distinct Gaussians 
stays positive definite but becomes severely ill-conditioned. The self-energy term is therefore a hardness offset derived 
from the width $w$, not the self-energy of the charge density that the off-diagonal kernel describes, which keeps the 
type's own width. The offset bounds the effective screening length from below, and the conductor limit of the model is a 
finite Thomas--Fermi length rather than a perfect conductor. Taking $w$ equal to the type's own width would make the 
functional that of one charge density, but it leaves the operator ill-conditioned and staggers the layer charges of a 
slab. The bound holds up to the discretization error of the particle mesh, which we measure below at 0.06 to 0.11~eV on a gold slab.

% =====================================================================
\section{Long-range electrostatics inside the matrix-vector product}
\label{sec:longrange}
% =====================================================================

Placing the reciprocal term inside the operator avoids an outer self-consistent loop. We carry the real-space Ewald-split short range in the 
sparse operator and add the reciprocal-space per-atom potential inside the matrix-vector product, so that one projected solve returns 
fully long-range charges with no self-consistent-field outer loop. The grid-dependent reciprocal self-interaction of each charge 
is removed from the operator diagonal, which then carries the on-site hardness and the interaction of the charge with its own 
periodic images. By default the term is calibrated from sixteen deterministic probe pairs, those of the lowest global tags, 
and re-measured only when the mesh or the Ewald parameter changes. It may instead be pinned to a value, or computed exactly for 
each atom from its interpolation stencil and one transform of the Green's function (\code{fix\_modify <id> recip\_self peratom}), 
which requires no calibration and does not drift under a barostat.

Evaluating the reciprocal term on a mesh within an iterative charge-equilibration solve is established practice.\cite{AhrensIwers2021,AhrensIwers2022,Zeng2021,Gubler2024,Savvidi2025} 
What \samQEq{} introduces is not the mesh but what the projector enforces, and how many conditions it carries at once: one constraint per fragment from the molecular connectivity. 
This ensures that the reciprocal evaluation sits inside a solve, carrying as many simultaneous constraints as there are molecules, each reimposed on every Krylov iterate. 
That is what makes the insulating limit of Section~\ref{sec:constraint} a run-time mode rather than a different code, and it is the reason a published 
bare-intermolecular model ports without refitting (Section~\ref{sec:port}).

A second difference is the kernel. Gaussian charge densities are the natural choice on a mesh, and Ref.~\citep{Savvidi2025} accordingly replaces Slater-orbital 
shielding with Gaussians whose widths are fitted, per element, to reproduce the Slater repulsion. We instead carry each kernel of Table~\ref{tab:kernels} as 
given on the long-range path: the exact Slater integral is tabulated and used unmodified, and the Gaussian carries the widths of the model that defines it. 
The splitting is the same for every kernel,
\begin{equation}
J_{ij}(r) \;=\; \underbrace{\left[ J_{ij}(r) - \frac{\mathrm{erf}(g_{\rm E} r)}{r} \right]}_{\text{real space},\; r < r_c}
\;+\; \underbrace{\frac{\mathrm{erf}(g_{\rm E} r)}{r}}_{\text{particle mesh}} ,
\label{eq:split}
\end{equation}
with $g_{\rm E}$ the Ewald splitting parameter and $r_c$ the solve cutoff. The mesh carries the point-charge term only and the kernel enters through the 
real-space bracket, which reduces to $\mathrm{erfc}(g_{\rm E} r)/r$ where $J_{ij}$ has reached $1/r$, and the cutoff must lie beyond that point. 
For the Gaussian kernel, Eq.~\eqref{eq:gauss-fourier} shows the equivalent reciprocal-space form. No kernel is refitted for mesh efficiency, 
and all three are exercised under full Ewald.

\samQEq{} provides two non-equivalent long-range modes. Shielding only intramolecular pairs and leaving intermolecular pairs bare reproduces the
structure of published bare-QEq models, and is the mode in which such a model ports faithfully. Shielding all pairs instead moves the solution away from the polarization catastrophe (the point at which the smallest eigenvalue of the constrained operator reaches zero), and supports stable 
energy-conserving dynamics. It does, however, move the model away from the bare-QEq reference. We therefore recommend the all-pairs mode for new parameterizations. 
The difference between the two modes is structural, not a fitting residual, as measured in Section~\ref{sec:port}.

% =====================================================================
\section{Coupling to other potentials}
\label{sec:coupling}
% =====================================================================

\samQEq{} is a layer independent of the potential: it solves only for the charges, 
the companion pair style adds the shielding correction $\Jij(r) - 1/r$ to the forces, and the long-range sum 
is standard particle-mesh, with an in-package routine supplying the per-atom reciprocal potential. 
Any other LAMMPS pair style is combined with it through \kw{hybrid/overlay}. It ships its own copy of the ACKS2 
augmented saddle-point solver, so it depends only on standard LAMMPS packages.

\subsection{Fixed-charge force fields}
\label{sec:fixedcharge}

A fixed-charge force field supplies the dispersion, repulsion and bonded terms, and \samQEq{} replaces its static charges by solved ones. 
Species whose charge should stay fixed, such as an ion held at an assigned charge state or a phase whose charges are not to be equilibrated, 
are left outside the solve group. The 13,216-atom cell of Section~\ref{sec:performance} composes the fix with an embedded-atom potential 
and a fixed-charge force field in this way. 

\subsection{Reactive force fields}
\label{sec:reactive}

A reactive force field carries its own charge-equilibration step, and the step can be delegated: \samQEq{} supplies the charges to the LAMMPS 
ReaxFF pair style with that style's own charge step disabled, using ReaxFF's electronegativities, hardnesses and shielding parameters under 
the same taper and global neutrality. Section~\ref{sec:reaxff} demonstrates this coupling.

\subsection{Machine-learned potentials}
\label{sec:mlip}

The same layering applies to a machine-learned potential without long-range electrostatics, in two ways. In the first, 
the electronegativities, hardnesses and widths are fixed per element, the machine-learned potential is trained on the 
difference between the reference energies and forces and those of \samQEq{}, and the two are summed 
at run time through \kw{hybrid/overlay}. In the second, the potential predicts $\chi_i$ from the local environment 
and \samQEq{} solves Eq.~\eqref{eq:solve} for the charges, which is the structure of fourth-generation potentials.\cite{Ko2021} 
Because the projector depends on neither 
$\chi$ nor the positions, the constrained minimum $E^\ast$ satisfies $\partial E^\ast/\partial\chi_i = q_i$, 
and the forces require no derivative of the charges,
\begin{equation}
\mathbf{F}_k \;=\; -\left.\frac{\partial E}{\partial \mathbf{R}_k}\right|_{\qvec}
\;-\; \sum_i q_i \frac{\partial \chi_i}{\partial \mathbf{R}_k} ,
\label{eq:mlip-force}
\end{equation}
where $\mathbf{R}_k$ is the position of atom $k$. The second term is a single backward pass through the potential with the solved charges as its input.
With the Gaussian kernel the reciprocal-space energy of the second mode is that of Latent Ewald Summation,\cite{Cheng2025} with the latent variable replaced 
by an equilibrated charge.\cite{Zhong2025,Ko2021} This coupling will be developed formally in a future work and is not demonstrated here.

% =====================================================================
\section{Code architecture}
\label{sec:implementation}
% =====================================================================

\subsection{Design goals}
\label{sec:design}

\begin{figure*}[t]
\centering
\includegraphics[width=\textwidth]{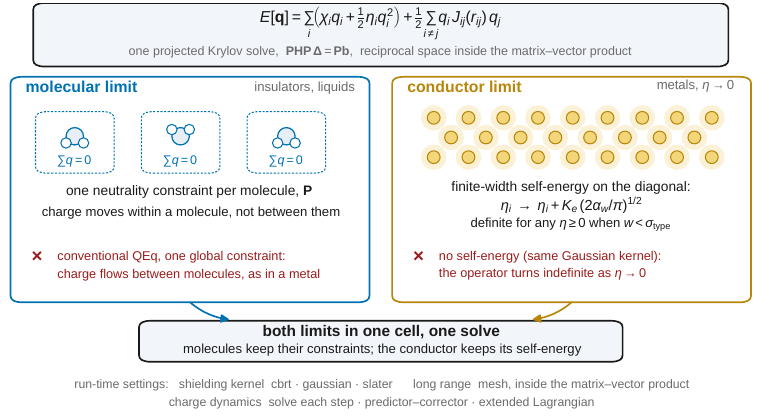}
\caption{One functional, two limits. Top, the energy functional, written for $\qzeroi{i}=0$, and the
single projected solve that minimizes it, with the reciprocal-space term inside the matrix-vector product
(Section~\ref{sec:longrange}). Left, the molecular limit: one neutrality constraint per molecule, imposed
by the projector (Section~\ref{sec:constraint}); under the single global constraint of conventional charge
equilibration, charge flows between molecules as in a metal. Right, the conductor limit: the finite-width
self-energy of a Gaussian charge on the diagonal keeps the conductor's part of the operator positive definite for any hardness
(Section~\ref{sec:conductor}); without the self-energy the operator turns indefinite as the hardness approaches
zero. Bottom, both limits in one cell and one solve (Section~\ref{sec:twolimits}), and the remaining
run-time settings.}
\label{fig:arch}
\end{figure*}

We now formally state the three requirements that shaped the implementation. The first is that one solve be well posed in both limits, 
for molecules and for conductors, and in a cell that holds both (Figure~\ref{fig:arch}): the constraint and the self-energy that make 
each limit well posed are properties of the atoms, not of the code path. The second is that the solver run beside any potential 
(Section~\ref{sec:coupling}), and the third that it impose no outer self-consistent loop (Section~\ref{sec:longrange}). 
The remaining choices on which published models differ (the shielding kernel, the long-range treatment and the charge dynamics) 
are run-time settings against the same solver, so that changing a model does not change the code.

\subsection{Matrix-free assembly}
\label{sec:matfree}

\begin{figure*}[t]
\centering
\includegraphics[width=\textwidth]{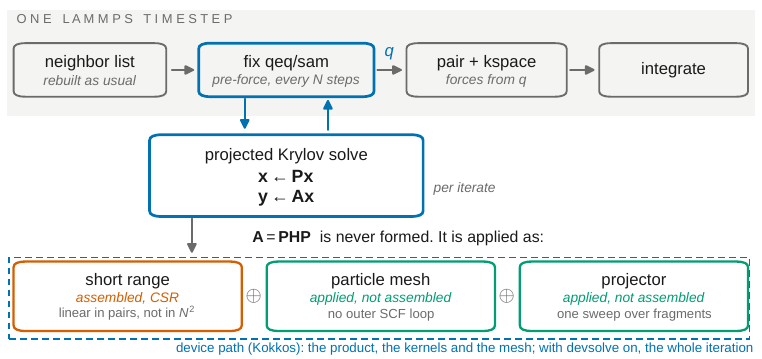}
\caption{Where the solve sits, and what the operator is made of. Top, one timestep: the fix runs
before the forces and hands the pair and long-range styles a charge set. Middle, the solve is a
projected Krylov iteration whose only argument is the action of the operator on the current iterate.
Bottom, that action is assembled in one part and applied in two: the short-range block is stored in
compressed sparse row form, while the particle-mesh reciprocal (Section~\ref{sec:longrange}) and the
per-fragment projector (Section~\ref{sec:constraint}) are evaluated on the fly, so the constrained
operator $\Amat = \Pmat\Hmat\Pmat$ is never formed. The dashed band marks what the Kokkos device path
replaces (Section~\ref{sec:device}). Schematic.}
\label{fig:callstruct}
\end{figure*}

\samQEq{} is matrix-free: the Krylov solver receives the action of the operator on a vector, never the matrix itself
(Figure~\ref{fig:callstruct}). That choice sets the memory footprint, the cost of 
adding a model, and what a device port has to replace. Only the short-range block is assembled, on the default route. 
Once per solve, \samQEq{} walks the neighbor list and stores
the shielded Coulomb couplings inside the shielding cutoff in compressed sparse row form, as a row offset, a neighbor count, 
a column index list and the values. Storage is therefore linear in the number of pairs inside that cutoff rather than quadratic in the number of atoms.

One advantage of this implementation is that the two terms that would dominate a dense assembly are never built. 
The reciprocal-space contribution is applied inside the matrix-vector product as a particle-mesh transform 
(Section~\ref{sec:longrange}), so the full long-range block, which couples every atom to every other, exists only as 
the action of a transform on the current iterate. The projector is likewise applied rather than formed: 
at each Krylov iteration the vector is swept once over fragments and its per-fragment
mean removed, at a cost linear in the number of atoms, so the rank-$n_{\rm f}$ constraint, with $n_{\rm f}$ 
the number of fragments, never appears as stored entries either. The constrained operator $\Amat = \Pmat\Hmat\Pmat$, 
whose spectrum decides whether the solve is a minimization at all (Section~\ref{sec:definite}), consequently has no 
representation anywhere in the code. It exists only as the composition of the operations the Krylov method calls.

The optional block incomplete-LU preconditioner of Section~\ref{sec:performance} is a deliberate exception: every rank gathers the 
global augmented system, carries the two global constraints of the augmented formulation as bordered Schur rows, 
and factors it, so that the preconditioner is independent of the domain decomposition by construction. This replicated 
factorization makes the solve practical on near-singular metal systems but limits it to thousands of atoms rather than millions.

Three consequences naturally follow. First, a new shielding kernel or a new long-range route is a change 
to the matrix-vector product and to nothing else, which makes the kernel and the long-range treatment 
run-time settings rather than code paths. Second, the device port has one object to replace rather than 
a solver to rewrite (Section~\ref{sec:device}). Third, a spectrum cannot be read from the running code 
by inspection. The spectra of Section~\ref{sec:definite} therefore come from the operator rebuilt outside 
the simulation, or recovered from the code's own matrix-vector product, and checked against the charges 
the code solves (Section~\ref{SI-sec:dense-native} of the Supplementary Material).

\subsection{Run-time controls}
\label{sec:controls}

The constraint, the self-energy of the conductor types and the settings of Section~\ref{sec:design} are 
selected through \code{fix\_modify} keywords, so that changing a model is an edit to an input deck rather 
than to the code. The charge model of the water benchmark takes five commands beyond those of a fixed-charge deck:

\noindent\begin{minipage}{\linewidth}
\begin{lstlisting}[language=LAMMPS]
pair_style   hybrid/overlay lj/cut 10.0 coul/long 10.0 coul/shield/intra 10.0 all
pair_coeff   * * coul/shield/intra 1.11
kspace_style pppm/samqeq 1.0e-6
# fix ID group qeq/sam Nevery cutlo cuthi tol paramfile
fix          chg fq qeq/sam 1 0.0 10.0 1.0e-5 spcfq_expt.param
fix_modify   chg recip_self peratom   # exact per-atom grid self-term
\end{lstlisting}
\end{minipage}

\noindent The listing shows the recommended settings. The water-benchmark decks behind 
Sections~\ref{sec:stability} to \ref{sec:performance} used a 5~\AA{} shielding cutoff 
and a pinned self-term, except the energy-conservation deck of Table~\ref{tab:propagators}. 
That table and Section~\ref{sec:limitations} measure what each choice costs.
The pair style must use the same kernel, kernel parameters and pair scope as the charge solve. 
The initialization check of Section~\ref{sec:checks} enforces this. Its cutoff must be close 
enough to the solve cutoff that the shielding correction is negligible where the shorter one stops, 
and energy conservation requires matching the two cutoffs more closely than that check 
demands (Section~\ref{sec:performance}). The final \kw{all} is the scope, which must match 
the sixth field of the parameter-file header (\code{lr\_ewald}) and so differs between decks 
that shield all pairs and decks that shield intramolecular pairs only. Table~\ref{tab:controls} lists the
principal controls; the package README gives the complete keyword and parameter-file reference.

The parameter file carries a header line and one line per atom type. The header is \code{gamma\_align kappa\_bond r\_ov r\_loc lr\_alpha lr\_ewald lr\_ridge}. The field 
\code{lr\_ewald} selects the long-range path: 0 evaluates the operator in real space alone; 1 and 2 place the reciprocal-space contribution inside the matrix-vector product, 
shielding intramolecular pairs only or all pairs (Section~\ref{sec:longrange}). \code{kappa\_bond} and \code{r\_ov} are the amplitude and length of the bond-softness term
$\kappa\exp(-r^2/r_{\rm ov}^2)$ within a fragment, \code{lr\_alpha} is the damping parameter of the damped-shifted-force (DSF) real-space path, and \code{lr\_ridge} is an optional diagonal ridge 
in the units of $\eta$; \code{gamma\_align} (which must be positive) and \code{r\_loc} are reserved and enter no solve in this
release. Each type line is \code{type} $\chi$ $\eta$ $\gamma$ $q^0$ $\varepsilon_{\rm HOMO}$ $\varepsilon_{\rm LUMO}$, where the
last two are likewise reserved, with two optional trailing columns carrying cubic and quartic on-site coefficients, 
which default to zero and recover the parabola of Eq.~\eqref{eq:functional}. The fourth column is not a single physical quantity: it is an inverse shielding length under 
\code{cbrt}, a Gaussian radius $R_c$ under \code{gaussian}, and a Slater exponent $\zeta$ under \code{slater} (Table~\ref{tab:kernels}).

\begin{table}[h]
\caption{Run-time solver controls, issued as
\code{fix\_modify <id> <keyword> ...}. $\lmin$ is the smallest eigenvalue of
the projected charge operator, estimated as in
Section~\ref{SI-sec:lanczos} of the
Supplementary Material; \code{<cap>} is a charge magnitude in units of
the elementary charge.}
\label{tab:controls}
\begin{tabular}{p{0.36\textwidth}p{0.54\textwidth}}
\toprule
Control & Effect \\
\midrule
\code{solver cg} $\vert$ \code{minres [<cap>]} &
Selects the long-range solver. The optional cap enables a bounded early termination, a heuristic safeguard, when the iteration stalls
(Section~\ref{SI-sec:minres} of the Supplementary Material). The default is \code{cg}, with an automatic
MINRES re-solve when a solved charge exceeds a threshold. \\
\code{solver cg nofallback} & Disables the automatic fallback. \\
\code{solver warmstart on} &
Seeds the solve from the previously converged charges, which reduces the
iteration count by about a quarter on the water benchmark (Table~\ref{tab:propagators}). \\
\code{ridge eig <floor> [<m>] [<every>]} &
Ridge that varies continuously with the configuration, set from an $m$-step Lanczos estimate of $\lmin$,
recomputed every \code{<every>} steps (piecewise constant between updates when \code{<every>} $> 1$). The ridge regularizes the solve and is not part of the reported
energy, so forces are not conservative while it is active. \\
\code{shield cbrt} $\vert$ \code{gaussian} $\vert$ \code{slater} &
Off-diagonal kernel, Table~\ref{tab:kernels}. \\
\code{shieldpair <i> <j> <R>} &
Overrides the Gaussian radius of one type pair, Eq.~\eqref{eq:betaij}. \\
\code{gself on [width <w>] [types ...]} &
Finite-width self-energy of Eq.~\eqref{eq:gself} on the listed types (default: every type, $w = 0.5$~\AA{}; list the conductor types, since on a molecular site it double-counts the hardness). \\
\code{xl <mass> <damp>} &
Extended-Lagrangian charge dynamics. \\
\code{aspc <n>} &
Time-reversible predictor with $n$ corrector iterations. \\
\code{spikeguard <q> <k>} $\vert$ \code{off} &
Soft projected restoring force on charges above $q$, for the
extended-Lagrangian path. On by default, with $q = 2\,e$ and $k = 10$~eV\,$e^{-2}$. \\
\code{precond ilu [<tol>]} &
Block incomplete-LU preconditioner of the augmented system, drop tolerance $10^{-3}$ by default. \\
\code{recip\_probes <K>} &
Number of probe pairs in the calibration of the reciprocal self-term, 16 by default; 0 selects
the single-pair estimator. \\
\code{bondsoft <kappa> <bcut>} &
ACKS2 bond-softness kernel with scale \code{kappa} and cutoff \code{bcut}; selects MINRES on the
short-ranged Gaussian operator. \\
\bottomrule
\end{tabular}
\end{table}

\subsection{Charge dynamics}
\label{sec:chargedyn}

The charges may be obtained in three ways, and the full solve may also be seeded from the
previous step (Table~\ref{tab:propagators}). The default solves Eq.~\eqref{eq:solve} each step by a projected conjugate gradient. An extended-Lagrangian option propagates 
the charges as fictitious-mass dynamical variables by velocity Verlet, as in the original fluctuating-charge models, avoiding the per-step solve.\cite{Rick1994} A time-reversible always-stable predictor--corrector (ASPC) predicts the charges from history and 
runs a small fixed number of corrector iterations.\cite{Kolafa2004} In ASPC, each corrected step is accepted only if it satisfies a residual criterion; otherwise the full 
solve is used. Section~\ref{sec:performance} gives the cost and the energy conservation of each on one system.

\subsection{Units}
\label{sec:units}

\samQEq{} supports two unit systems, \kw{metal} (eV) and \kw{real} (kcal\,mol$^{-1}$), under three rules. First, every quantity typed by the user (parameter-file columns and 
run-time modifier arguments alike) is in the deck's native energy units and is never pre-converted. Second, internal hard-coded constants are anchored in electronvolts and 
converted once through a single scale factor, so that they mean the same physical value in either system. Third, solver tolerances are rescaled internally, so that
the same numerical tolerance means the same physical convergence in both systems.

\subsection{Device kernels}
\label{sec:device}

The object the device has to replace is the matrix-vector product of Section~\ref{sec:matfree}, not the solver. Under the LAMMPS Kokkos package, selected with the standard suffix, 
the sparse product, the three shielding kernels and the particle-mesh reciprocal all have device implementations, and with \code{fix\_modify <id> devsolve on} the projected 
conjugate-gradient iteration runs on the device as well, including under ASPC. The iterate, the residual and the search direction are device arrays, and with the device mesh style 
the reciprocal term stays on the device too. One per-iteration host transfer remains, the ghost exchange of the trial charges, which LAMMPS routes through host memory for a fix. 

Two limitations apply. First, not every route is ported. Routes that run on the host solve include the quartic on-site
term, the per-atom ridge, the bond-softness operator and the intramolecular-only shielding mode of Section~\ref{sec:longrange}.
Per-pair Gaussian radii, ion-damped kernels, the exact per-atom self-term, a molecule-free long-range solve and slab geometries (\kw{kspace\_modify slab}) are also not
ported. The device styles refuse these at initialization, and such decks run with the host styles; the conductor slabs of this paper are therefore host-only. We recommend checking any deck that requests the device solve against the host
result. Second, halo exchange goes through GPU-aware MPI, which is not an optimization but a requirement of practical throughput: staging the same halos through host 
memory costs an order of magnitude on four devices (Section~\ref{sec:performance}).

We validated the device port against forces in addition to energies and charges. The potential energy and the solved charges are contractions that a defect in a field-carrying 
ghost exchange can leave numerically exact, so a bitwise comparison of either passes while the force on an atom near a subdomain face is wrong. We therefore require the device path to reproduce the host forces at fixed charges to $10^{-11}$~eV~\AA$^{-1}$ on one
and two GPUs, and the host charges to $10^{-11}\,e$ after ten steps of the benchmark of Section~\ref{sec:performance}
(Section~\ref{SI-sec:si-gpu} of the Supplementary Material).

% =====================================================================
\section{Numerical stability}
\label{sec:stability}
% =====================================================================

\subsection{Well-posedness of the solve}
\label{sec:definite}

The constrained operator is symmetric by construction, and whether it is also positive definite decides whether the solve is a minimization at all. This question is intrinsic
to QEq models: Savvidi and co-workers mapped the smallest eigenvalue against hardness and shielding width and noted that a solve on an indefinite matrix may return a saddle point 
or fail to converge.\cite{Savvidi2025} For the Gaussian kernel, Eq.~\eqref{eq:gram} gives a rigorous sufficient condition, up to mesh error; for the cube-root and Slater kernels
we use a weaker sufficient condition.

For a molecular system a sufficient condition is that every site's hardness exceed the total coupling that site carries,
\begin{equation}
\eta_i \;>\; \sum_{j \ne i} \left| \Jij(r_{ij}) \right| \quad \text{for every } i ,
\label{eq:bound-ineq}
\end{equation}
which is row-wise diagonal dominance of $\Hmat$ and passes to the constrained operator. It applies only to the truncated real-space operator: the reciprocal-space sum is not 
absolutely convergent, so for the full operator the condition is a heuristic. Because the kernel is monotone in $r$, each term is bounded by its contact value, so 
Eq.~\eqref{eq:bound-ineq} follows from $\Jij(0) < \min_i \eta_i/n_i$, with $n_i$ the number of neighbors of $i$ inside the cutoff. On a two-site fragment, where $n_i = 1$, 
that reduces to the per-pair statement $\Jij(0) < \min_i \eta_i$ in which the condition is usually quoted. On an extended system the per-pair statement is not sufficient: the sum runs over all neighbors, 
so a configuration whose lowest constrained mode alternates in sign between near neighbors can satisfy the per-pair bound and still carry a negative eigenvalue. For the Gaussian 
kernel no such collective analysis is needed: the gold slab's smallest eigenvalue, 3.47~eV below its hardness, is the contact value of its own kernel, Eq.~\eqref{eq:gram}, to within the mesh residual.

Equation~\eqref{eq:bound-ineq} is a condition on the parameters and the geometry rather than on the kernel form, and it is sufficient rather than necessary; the water benchmark 
of Section~\ref{SI-sec:water} is a case that violates it and is definite anyway. It violates the condition because its cube-root kernel reaches $\Jij(0) = 15.98$~eV against $\eta_{\rm O} = 15.11$~eV,
and it stays definite because its rigid geometry holds the intramolecular couplings low enough that the constrained operator of an isolated molecule has eigenvalues of 2.26 and 7.79~eV. Direct diagonalization
of the constrained operator of the 512-molecule liquid, with the reciprocal-space term included as the solve applies it, returns $\lmin = 1.36$~eV and a condition number of 8.0, 
with no negative eigenvalue (Section~\ref{SI-sec:dense-native} of the Supplementary Material). With the same kernel, a Slater exponent read as an inverse shielding length makes 
the operator of a single molecule indefinite, and the solve then returns a sign-inverted charge distribution (Section~\ref{sec:port}).

For a conductor the dominance bound cannot hold, since $\eta \to 0$, and the sufficient condition is instead Eq.~\eqref{eq:gram} with the self-energy of Eq.~\eqref{eq:gself}. 
Equation~\eqref{eq:gram} is itself sufficient rather than necessary. The M site of the water model of Section~\ref{sec:twolimits} has a hardness below its own contact value, 
16.11 against 17.53~eV, so the bound does not necessarily validate the model. The dense spectrum of the interface operator is nonetheless positive, with $\lmin = 1.68$~eV under per-molecule 
constraints and 1.53~eV under the global one. The bound is a worst case over single sites, and under either constraint no admissible charge displacement is confined to one site.

\subsection{Solvers}
\label{sec:minres}

The default solver is a projected conjugate gradient. The package also provides an augmented saddle-point formulation. Because the augmented matrix is indefinite, that path 
uses BiCGStab with a restart that re-anchors the shadow residual when it becomes orthogonal to the residual, a breakdown that otherwise returns incorrect charges.\cite{vanderVorst1992} 
A projected MINRES solve\cite{Paige1975} is available for indefinite operators. On a 216-molecule fluctuating-charge
water box it costs the same as preconditioned conjugate gradient (18 matrix-vector products each at a tolerance of
$10^{-7}$), because the matrix-vector product dominates the solve (Section~\ref{sec:performance}). The near-uniform
hardness diagonal makes diagonal preconditioning worth little.

\subsection{Consistency checks at initialization}
\label{sec:checks}

A converged residual on a neutral charge set does not ensure that the intended problem was solved. The residual is formed against the operator in use, and the projector 
holds the charges neutral whatever operator produced them, so the two quantities a user is most likely to inspect are both blind to a substitution of the operator itself. 
\samQEq{} therefore provides the following explicit checks at initialization.

When only part of the system is solved, \samQEq{} tests group membership for both atoms of every matrix element, so that no element couples a solved atom to an atom whose 
entries were never assembled. It asserts that the companion pair style and the fix carry the same kernel, the same parameters and the same set of pairs, and that the 
pair style has a term for every pair of solved types (an unassigned pair is refused, or flagged when the shielding is intramolecular only): under LAMMPS's \kw{hybrid/overlay}, coefficients given per type are not mixed across type pairs that another sub-style 
already covers, so the cross pairs would enter the solve and not the forces. It compares the two cutoffs by magnitude, asking whether the shielding correction is still 
appreciable where one side stops applying it. It requires its calibrated constants to be reproducible across domain decompositions and checks each recalibration against 
the first at the same mesh; the exact per-atom self-term of Section~\ref{sec:longrange} removes that calibration altogether. A self-term pinned by the user belongs to
one Ewald splitting, and \samQEq{} warns when it departs by more than 2\% from the converged-mesh limit $2g_{\rm E}/\sqrt{\pi}$ of the current one: on the water benchmark, 
a pin carried unchanged from a 10 to a 12~\AA{} cutoff is 19\% off and drives the liquid close to the polarization catastrophe. The convergence tolerance admits no such check: 
whether a tolerance is tight enough depends on the property of interest, not on the solve. We bound it by measurement instead (Section~\ref{SI-sec:si-tolscan} of the Supplementary Material).

% =====================================================================
\section{Verification and validation}
\label{sec:validation}
% =====================================================================

\subsection{Molecules and a conductor in one solve}
\label{sec:twolimits}

The central thesis of this work is that one solve can hold a molecular liquid and a metal at once, each in its own limit. We test it on the 240-atom Au(111) slab 
carrying a film of 81 water molecules on its upper face, with no applied field, so that the metal responds only to the water. The water
is described by the TIP4P-FQ\cite{Rick1994} force field with the Gaussian kernel, with the two radii fitted to reproduce the model's own intramolecular Coulomb integrals at its rigid geometry 
(Section~\ref{SI-sec:twolimits} of the Supplementary Material). Two settings distinguish the four models: (i) the constraint, either one global neutrality condition over
metal and water or one per water molecule plus one for the slab; and (ii) the finite-width self-energy of
Eq.~\eqref{eq:gself} on the gold, on or off. The per-molecule constraint is a modeling choice: it forbids charge transfer between the water and the metal, and with it 
charge-transfer chemisorption and interfacial redox. We scan the gold hardness from 5.172~eV, which corresponds through $\eta = 4\pi K_e l_{\rm TF}^2 d$ to a Thomas--Fermi 
length of 0.70~\AA{} at the atomic density of gold, down to 0.05~eV. 

\begin{figure*}[t]
\centering
\includegraphics[width=\textwidth]{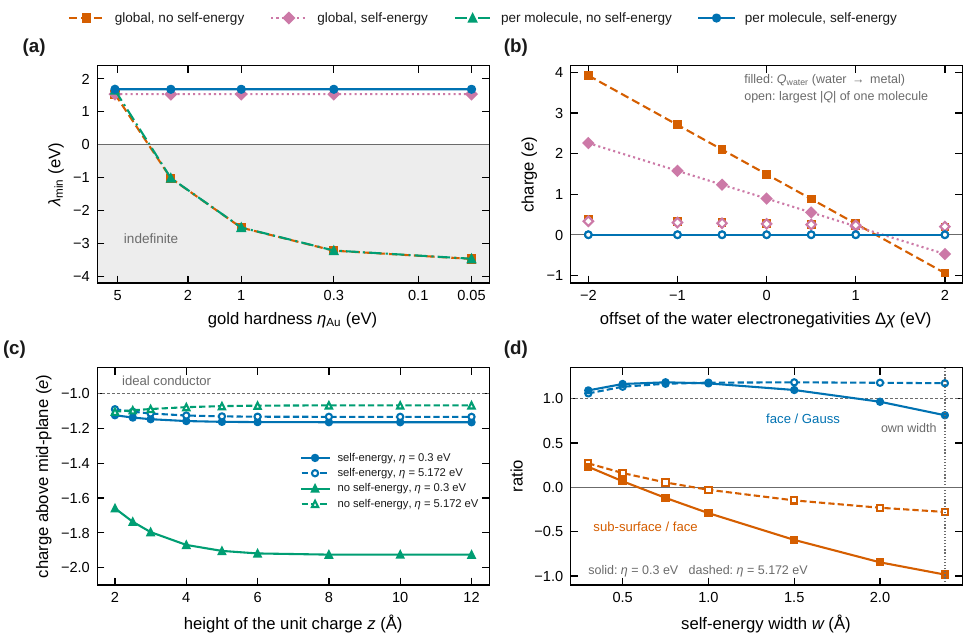}
\caption{Molecules and a conductor in one solve. (a) Smallest eigenvalue of the constrained charge
operator of an Au(111) slab carrying 81 water molecules, by dense diagonalization, against the gold
hardness (logarithmic, decreasing to the right), for the four combinations of constraint and self-energy; shading marks an indefinite operator.
(b) At the Thomas--Fermi hardness, the net charge moved from the water into the metal (filled) and the
largest net charge on a single molecule (open), against a common offset of the water electronegativities;
the two per-molecule arms coincide at zero and are invariant to the offset. (c) Charge above the mid-plane
of a 2160-atom slab, from the Gaussian charge densities, opposite a unit point charge at height $z$, with
and without the self-energy, at the Thomas--Fermi hardness (open) and at 0.3~eV (filled); the dotted line
is the ideal conductor. (d) The self-energy width: face charge in a field of 0.5~V\,\AA$^{-1}$ relative to the
Gauss limit, and the sub-surface layer relative to the face, whose sign change marks the staggered mode;
the dotted line is the width of gold's own Gaussian.}
\label{fig:conductor}
\end{figure*}

With the self-energy the effective diagonal of Eq.~\eqref{eq:gram} is $\eta + 12.83$~eV, so the scan moves the equivalent screening length only from 1.30 to 1.10~\AA{}. 
The lower end is the conductor limit of the model, not a vanishing screening length. Each point is a single solve on one fixed configuration, and $\lmin$ is from dense 
diagonalization of the operator the code applies, recovered column by column from its own matrix-vector product; that operator reproduces the solved charges to $2.1\times10^{-7}\,e$.
No eigenvalue is negative at any hardness. Under per-molecule constraints the smallest eigenvalue is 1.68~eV throughout, set by the water; the metal's modes lie above 13~eV, and the solve 
converges in 31 to 35 iterations at every hardness. Conversely, without the self-energy the operator follows the gold's bound, $\lmin \approx \eta - J_{\rm AuAu}(0)$, 
and turns indefinite once the hardness falls below 3.5~eV, with up to 221 negative eigenvalues, under either constraint (Figure~\ref{fig:conductor}a). The corresponding solve 
takes 150 to 600 iterations and returns layer charges of up to 30~$e$. 

The global constraint fails differently, and the failure does not depend on the hardness. The electronegativities of a QEq water model carry an arbitrary common offset, 
which a per-molecule constraint cancels and a global one does not. For the global constraint, shifting the water's electronegativities by $\Delta\chi$ moves the net charge between 
water and metal linearly, from $+3.9\,e$ to $-0.9\,e$ over $\pm2$~eV (Figure~\ref{fig:conductor}b). The magnitude of the transfer therefore carries no physical information. 
What does not depend on the offset is the charge on individual molecules. Under the global constraint the largest net charge on a single molecule never falls below 0.19~$e$, 
even where the net transfer vanishes, because the constraint lets charge flow between molecules as in a metal. Under per-molecule constraints every molecule is neutral by construction, and 
the charges are invariant to the offset to $2\times10^{-11}\,e$; the mean water dipole equals that of the same film without the metal to 0.1\%, and the slab responds with a 
smooth image charge beneath the film. The demonstration is one interface in one configuration, and it establishes that both limits are well posed together; it does not test the physics of the water--metal coupling, which needs a fitted interface potential and a mobile metal.

A conductor must screen a charge at any distance. We measure the screening on a model system: a 2160-atom slab (the 240-atom slab
replicated three times in each lateral direction; Section~\ref{SI-sec:twolimits} of the Supplementary Material), as the charge
above the slab's mid-plane opposite a unit point charge. The charge is integrated from the Gaussian charge densities of the
kernel rather than taken from site charges, whose densities overlap adjacent layers. With the self-energy activated, it is $-1.13$ and $-1.16\,e$ at the two hardnesses and does not 
change with the height of the charge beyond 4~\AA{} (Figure~\ref{fig:conductor}c). The same definition gives 1.13 to 1.16 times the Gauss limit for the slab in the field of 
0.5~V\,\AA$^{-1}$. Both tests measure the laterally uniform response of the slab, and both read 1.13 to 1.16 times the ideal for the same reason. The gold Gaussian, with a standard deviation of 2.38~\AA{}, is wider than the equivalent Thomas--Fermi length of 1.1 to 1.3~\AA{}, so the induced charge does not decay monotonically into the metal but overshoots and oscillates, with a decay length of 2.9~\AA{} and a period of about 13~\AA{}. The mid-plane of a five-layer slab lies two layers below the face, inside the first overshoot, and the measure records it. On slabs of 13 and 17 layers the same model screens a unit charge to $-0.99\,e$ and a field to 0.99 to 1.00 times the Gauss limit, and a narrower gold Gaussian ($R_c = 0.5$~\AA{}) removes the excess on five layers as well. Without the self-energy term, the definite slab at the Thomas--Fermi hardness screens to $-1.07\,e$, and the 
indefinite one at 0.3~eV returns $-1.9\,e$. Beyond 4~\AA{} the interaction energy follows the exact image law of the laterally periodic cell to 23 and 27~meV with the self-energy, 
with the image plane 0.2 and 0.4~\AA{} above the top atoms; closer in, the charge's Gaussian overlaps those of the metal and the point-charge law no longer applies. Interestingly, 
the indefinite solve follows the law to 18~meV, meaning the energy cannot be used to discriminate between the two approaches.

The same film under dynamics separates the four combinations more sharply than any single solve. Here, we held the gold fixed,
kept the water rigid and thermostatted at 298~K, and ran each combination for 5~ps at a 0.2~fs step, at either hardness. With
the self-energy and per-molecule constraints, the largest charge on any site stayed at or below 1.33~$e$, under the 1.34~$e$ of the same 
film without the metal, and the eigenvalue ridge never engaged. Without the self-energy, at 0.3~eV, the per-molecule run raised no warning,
but the gold site charges drifted as the water moved, to a standard deviation of 0.63~$e$, and the largest charge, 1.88~$e$
after 5~ps, did not reach a steady value. The global 
constraint at the same hardness went near-critical within 0.3~ps, engaged the ridge at nearly every step thereafter, and moved up to 9.5~$e$ into the water, with single molecules 
carrying up to 1.0~$e$. At the Thomas--Fermi hardness the global constraint was stable, but single molecules carried up to 0.28~$e$ throughout.

The self-energy width is a parameter of the model, and the screening is sensitive to it (Figure~\ref{fig:conductor}d). For widths up to 0.5~\AA{} the sub-surface layer carries the sign of the face, 
a screening tail, and the face charge lies within 6 to 16\% of the Gauss limit. Between 0.75 and 1.0~\AA{} the sub-surface layer changes sign, and it stays
inverted at larger widths. At the width of gold's own Gaussian, 2.38~\AA{}, where the Coulomb block is only semidefinite, the layers alternate at about three times the Gauss limit 
at 0.3~eV, and the metal's softest mode falls below the water's. We use 0.5~\AA{} and recommend widths no larger than it in practice.

\subsection{Reproduction of a published model without refitting}
\label{sec:port}

The most direct test of the implementation is whether it reproduces a published model from that model's own parameters. Using the exact Slater kernel with the original Slater 
exponents of the TIP4P-FQ model of Rick and co-workers, in the intramolecular-shielding mode that matches the structure of their model, we obtain a gas-phase dipole of 1.8634~D against their
1.85~D and a liquid dipole of $2.630 \pm 0.004$~D
over seven of eight independent 100~ps trajectories against their 2.637~D.\cite{Rick1994} The eighth settles during equilibration into a less hydrogen-bonded network, with a dipole of 2.55~D that persists for the whole run. As in the original model, the port runs at a 0.25~fs step under a thermostat; we did not test it in the microcanonical ensemble. The tabulated kernel itself agrees with direct numerical 
quadrature of the reference integral to a maximum relative deviation of $5.2\times10^{-15}$.

To establish the sensitivity of the choice of shielding, we ran the same kernel and the same parameters in the all-pairs-shielded mode, which gives a liquid dipole of $2.340 \pm 0.001$~D 
over three independent 10~ps trajectories, an under-polarization of about 11\%. Both uncertainties are the standard error of the mean over independent trajectories, each trajectory contributing its time average. The difference comes from which pairs are shielded, not from the kernel or its parameters, and \samQEq{} 
therefore makes the pair scope a user setting. The pair style must be told to shield the same set of pairs as the solve; a deck that shields all pairs in the charge matrix while correcting 
only intramolecular pairs in the forces is an instance of the pair--solve mismatch that Section~\ref{sec:checks} guards against, and it shifts this
dipole by 0.12~D. Here, at a given configuration, the solved charges are unchanged, but the forces differ, and with them the sampled structure.

The effect of kernel choice itself can be isolated in the same way, with an even larger effect. Table~\ref{tab:kernelcmp} compares two runs of the gas-phase monomer that differ by a single 
substantive line of input, the kernel selection. Read as Slater exponents under the exact integral, the published parameters give the published dipole. Read as inverse shielding 
lengths under the cube-root kernel, the same numbers give a dipole of similar magnitude but a sign-inverted charge distribution: the M site (the off-atom charge site of the model), 
which carries the largest electronegativity at $\chi_{\rm M} = 7.47$~eV and must therefore be negative, comes out positive.

The inversion originates in the operator of the isolated molecule. Interpreted as $\gamma$, these exponents raise the cube-root kernel to $J(0)/\eta$ 
of 2.15, 2.75 and 1.60 for the H--M, M--M and H--H type pairs, and at the geometry of the model they make the constrained operator indefinite, with eigenvalues of $-2.18$ and $+5.98$~eV. 
The native model, whose contact value also exceeds its hardness, stays positive definite (Section~\ref{sec:definite}). The resulting stationary solution has an inverted charge 
distribution. Neither kernel is inferior: a Slater exponent and an inverse shielding length are parameters of different kernels, and reusing one number under the other is not a small 
approximation but an entirely different model (here a qualitatively wrong one). Separating the two requires that the kernel be a setting rather than a property of the code.

\begin{table}[t]
\caption{Two runs of the gas-phase monomer differing in a single setting, the
selection of the shielding kernel. The molecule, the parameter file and all
other settings are identical; the fourth column is read as a Slater
exponent in the first row and as an inverse shielding length in the second. The
reference dipole is 1.85~D.\cite{Rick1994} Charges are in units of the
elementary charge. Each run is one molecule in a cubic periodic cell, solved with the reciprocal-space
term inside the matrix-vector product and extrapolated to infinite dilution from cells of 100 and 120~\AA{}.
$E$ is the reported potential energy: the shielded pair and reciprocal-space energies and the on-site term
of Eq.~\eqref{eq:functional}. The inverted reading has
a \emph{positive} total energy, which the charge signs alone do not reveal. Both kernels are exercised by
a distributed two-deck example, \code{examples/kernel\_comparison}, in a 40~\AA{} cell.}
\label{tab:kernelcmp}
\begin{tabular}{lccccc}
\toprule
Kernel & $q_{\rm M}$ & $q_{\rm H}$ & $\mu$ (D) & Deviation & $E$ (eV) \\
\midrule
\code{slater} & $-0.8898$ & $+0.4449$ & 1.8634 & $+0.7\%$  & $-1.3213$ \\
\code{cbrt}   & $+0.9089$ & $-0.4544$ & 1.9033 & $+2.9\%$  & $+1.3497$ \\
\bottomrule
\end{tabular}
\end{table}

\subsection{Reproduction of the ReaxFF charge-equilibration step}
\label{sec:reaxff}

The charge equilibration of ReaxFF\cite{vanDuin2001,Aktulga2012} is a 
tapered QEq whose kernel is the cube-root form of Table~\ref{tab:kernels}. We therefore ran the LAMMPS ReaxFF pair style with its own charge step disabled and let \samQEq{} supply
the charges from ReaxFF's electronegativities, hardnesses and shielding parameters, under the same taper and global neutrality, with no parameter refitting. On the 3000-atom QEq water 
of the LAMMPS distribution\cite{Achtyl2015} and on a 21-atom RDX molecule, the charges agree with the native ReaxFF step to at most $1.2\times10^{-9}\,e$ and $5.5\times10^{-10}\,e$, 
respectively. The ReaxFF energy agrees to $8\times10^{-9}$ and $1\times10^{-11}$~kcal~mol$^{-1}$, and the forces to $7\times10^{-8}$ and $2\times10^{-8}$~kcal~mol$^{-1}$~\AA$^{-1}$, 
for the water and RDX, respectively. A 20-step thermostatted trajectory of the water tracks the native one to $5.5\times10^{-8}$~kcal~mol$^{-1}$ in potential energy. One caveat applies: 
we convert the parameters with ReaxFF's rounded Coulomb constant, 14.4~eV~\AA{}, rather than the exact one, and with the factor of two with which ReaxFF converts its tabulated hardness. 
The exact constant would move the charges by at most $9\times10^{-5}\,e$, and omitting the factor of two moves them by up to $3.1\,e$.

The reproduction ships as \code{examples/reaxff\_port}, with the reference and \samQEq{} decks for both systems, the script that converts the parameters and the reference output. 
It is an example rather than a regression case because it needs the LAMMPS ReaxFF package, which the rest of the package does not. The composition needs three settings besides 
the parameters: the fix's own on-site energy is excluded, because the ReaxFF pair style already carries it, and the particle-mesh and long-range pair terms that
the fix requires are present but switched off. Omitting any of them leaves the charges correct and the energy and forces wrong.

\subsection{Regression suite}
\label{sec:regression}

Numerical reproducibility across code changes is enforced by \ncases{} self-contained test cases, each run on a single processor, so that the result is bit-reproducible, and 
compared against committed per-atom charges at a relative tolerance of $10^{-9}$ and an absolute tolerance of $10^{-12}$. The suite exits non-zero on any failure and runs from 
a pre-commit hook, so a change that perturbs a converged charge cannot be committed without the perturbation being seen. The cases span the gas and liquid models in both 
long-range paths, all three shielding kernels, the extended-Lagrangian and ASPC propagators, solver set-up across successive runs, molecule-free global-neutrality solves, slab geometries, 
the gold slab of Section~\ref{sec:conductor} with its finite-width self-energy, the water film on gold under both constraints, the augmented solve with its preconditioner, and the exact per-atom grid self-term, with separate scripts in the package for multi-rank runs and for the device styles (Section~\ref{SI-sec:tests} of the Supplementary Material).

\section{Known limitations}
\label{sec:limitations}

The present release has the following limitations; the list is not exhaustive. 

\begin{itemize}[label=\scalebox{0.5}{$\blacksquare$}]

\item Constant-potential routes, in which the electrode charges are themselves prescribed rather than solved under a neutrality constraint, are the subject of future work and are 
not part of this release.

\item The physical validation reported here is on water, an Au(111) slab with and without a water film, and two ReaxFF systems. The accuracy of the framework is not yet demonstrated on an 
organic electrolyte, an ionic liquid or a metal/electrolyte interface, which are the natural next tests of transferability. The conductor mode is measured at one field strength, on Au(111) slabs of 3 to 17 layers, with the width scanned in Figure~\ref{fig:conductor}d, and the width is a parameter that the shipped sets do not supply in general; it must be derived
for each new electrode material. The two limits are shown together on one fixed configuration of one interface (Section~\ref{sec:twolimits}) and over 5~ps of dynamics with 
the metal held fixed; mobile copper electrodes run stably for 5~ps in the cell of Section~\ref{sec:performance}, but a production-length metal/electrolyte interface with fitted metal--electrolyte interactions is not yet validated in this mode, and the metal/electrolyte cell of Section~\ref{sec:performance} 
runs the conductor mode with mobile electrodes for 5~ps as a stability and throughput test, with an illustrative zinc--copper contact term; it is not a validated model of zinc deposition. No machine-learned potential is coupled here.

\item The default atomic-charge solve should not be used for the longitudinal response of long molecules: it makes their longitudinal polarizability grow super-linearly with chain length
rather than linearly. The remedy, the bond-softness kernel, requires two additional parameters. The values that restore the correct scaling give 0.62~\AA$^3$ per carbon, about a third of the polarizability 
increment of a CH$_2$ group, 1.84~\AA$^3$.\cite{Miller1990} We expect condensed-phase properties, where short-range intermolecular polarization dominates the response, to be much less sensitive; the water benchmark 
results reported here use the atomic-charge mode.

\item The device path does not strong-scale across nodes for the system sizes considered here. On one node it runs about twice as fast as a CPU node, and four nodes return only 1.2 to 1.3 
times one node, because roughly thirty latency-bound collectives per timestep sit inside the solve (Section~\ref{sec:performance}).

\item The initialization check of Section~\ref{sec:checks} compares the fix against the companion pair style only. When the Coulomb term of the forces is supplied by another pair style, 
as in the ReaxFF composition of Section~\ref{sec:reactive}, nothing verifies that its kernel matches the solve, and that composition also requires the fix's own energy and the 
standard long-range pair and mesh terms to be excluded by hand.

\item The extended-Lagrangian and ASPC charge propagators do not conserve energy for the water benchmark at the documented settings (Section~\ref{sec:performance}); a thermostat hides the failure. Use them only in thermostatted runs whose observables are checked against the full solve.

\item The MINRES path is unpreconditioned. That is adequate for a near-uniform hardness diagonal. It is not adequate for a dense-metal electrode, whose severely ill-conditioned operator 
requires the block incomplete-LU preconditioner of the augmented path (Section~\ref{sec:performance}).

\item The probe calibration of the reciprocal self-term has a validity window that narrows as the particle-mesh accuracy is tightened, because each probe must find a
partner within 2.5 mesh spacings for the Ewald complement to hold. At a particle-mesh accuracy of $10^{-7}$ the mesh spacing is small enough that a probe on an atom 
with a bond shorter than 1~\AA{} finds no partner. When the probes are requested explicitly, a probe without a partner is refused; otherwise the calibration falls back 
to the single-pair estimator, as it does for the gold slab of Section~\ref{sec:conductor}. The exact per-atom route has no such window, but it is host-only and requires 
a mesh of at least nine points per direction at fifth-order interpolation. The water benchmark is defined at its pinned self-term, 1.0\% below the converged-mesh limit
$2g_{\rm E}/\sqrt{\pi}$ of its Ewald splitting, and the liquid dipole depends on that choice: re-solved on the same 51 configurations at the same splitting, it is 2.5\% higher 
with the probe calibration and 3.5\% higher with the exact per-atom self-term, with a spread over configurations of 0.002~D. The benchmark parameters are fitted at the pinned
value and carry that choice. The benchmark lies close enough to the polarization catastrophe that a 1\% change in a diagonal term moves its dipole by 2.5 to 3.5\%. We therefore 
recommend the exact per-atom route wherever the mesh allows it. A pinned value also has to follow the splitting, which the particle-mesh style re-tunes at every \kw{run} from 
the current charges unless it is fixed with \kw{kspace\_modify}; the warning of Section~\ref{sec:checks} catches that case.

\end{itemize}

% =====================================================================
\section{Performance}
\label{sec:performance}
% =====================================================================

The CPU nodes of the Perlmutter system at the National Energy Research Scientific Computing Center carry two AMD EPYC 7763 processors (128 physical cores) and 512~GB of memory 
behind a single HPE Slingshot~11 interface. We benchmark a thermostatted 13,216-atom metal/electrolyte interface cell of $36.15 \times 36.15 \times 158.3$~\AA{}, in which 
two copper electrodes of 1,600 atoms in total bound 3,800 water molecules, 72 Zn$^{2+}$ and 144 Cl$^-$ ions. Each copper electrode is one fragment carrying the finite-width self-energy of Eq.~\eqref{eq:gself} ($w = 0.5$~\AA{}), so its charges are solved as a neutral conductor, and the copper moves: copper--copper interactions are the embedded-atom potential of Mendelev and King,\cite{Mendelev2013} and the electrodes are integrated and thermostatted with the electrolyte. Each zinc ion is a single-atom fragment held at $+2$ by the constraint, the water charges are solved with the Gaussian kernel, and the chloride ions stay outside the solve group; the zinc--copper short-range term is an illustrative Lennard-Jones potential, not a fitted one. ASPC advances the charges without the eigenvalue ridge, and SHAKE constrains the water at a 0.25~fs timestep. Over 5~ps at 298~K the electrodes stay crystalline (a copper mean-square displacement of 0.14~\AA$^2$) and neutral to $10^{-15}\,e$, the induced charge sits in the two layers facing the electrolyte, with site charges up to 0.25~$e$, and ASPC falls back to the full solve on 7 of 20,000 steps. On this cell the implementation sustains 8.5, 12.5 and 16.4 timesteps per second at 16, 32 and 64 processes, corresponding to 73\% and 66\% parallel efficiency at each of the two doublings of the process count (48\% from 16 to 64), within 3\% of the same cell with the copper held fixed and neutral atom by atom. All three counts fit within one node, so these are intra-node figures, and the loss in efficiency reflects
contention within the node rather than the network. At 64 processes the charge solve accounts for 26\% of wall time against 65\% for the pair interactions, and between 20 and 26\% at 16 to 128 processes, as the collectives inside the solve become latency-bound. The solve is the minority term because ASPC holds it to about two matrix-vector products 
per step; the iteration count sets that share, and at two products per step the choice of Krylov method has
little effect on throughput.

On the GPU nodes of Expanse at the San Diego Supercomputer Center, which carry four NVLink-connected NVIDIA V100 GPUs with 32~GB each, we further benchmark the water model of 
Section~\ref{sec:longrange}, replicated from 12,288 to 786,432 atoms under a Nos\'e--Hoover thermostat with the device solve, and the same deck on a full Expanse CPU node of two 
64-core AMD EPYC 7742 processors (Table~\ref{tab:gpu}; protocol in Section~\ref{SI-sec:si-gpu} of the Supplementary Material). A single V100 holds the largest box in 
under 24~GB and runs at 0.3--0.8 times the speed of the CPU node. Four V100s run 1.86 to 1.98 times as fast as the CPU node from 98,304 atoms upward. Both benchmarks are thermostatted, 
so where they use the predictor--corrector they measure its cost and not its energy conservation, which Table~\ref{tab:propagators} reports. Two choices decide those numbers. Solving the charges on the device 
is 4.4 times faster on one GPU than solving them on the host between device kernel calls, and exchanging halos through GPU-aware MPI raises the throughput tenfold over staging them 
through host memory at 331,776 atoms on four GPUs.

\begin{figure*}[t]
\centering
\includegraphics[width=\textwidth]{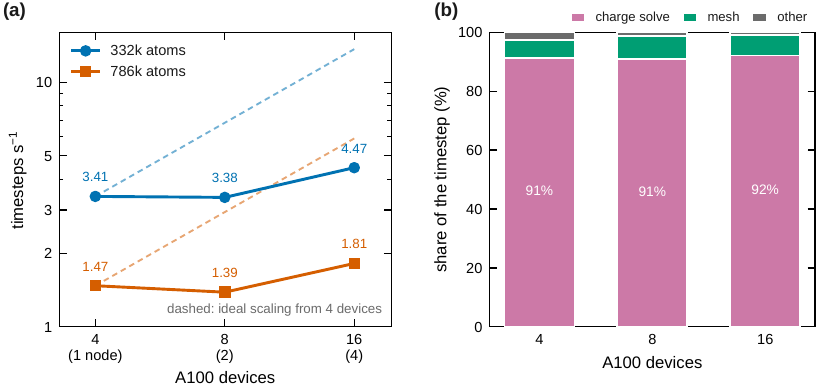}
\caption{Multi-node behavior of the device solve on Perlmutter A100 nodes, 200 timesteps with the full
solve at every step. (a) Measured throughput against device count at two system sizes; the dashed lines are
ideal scaling anchored on the single-node point. Two nodes give no measurable improvement over one, and four
give 1.2 to 1.3 times. (b) The share of the timestep by LAMMPS task category at 786,432 atoms. The charge solve is
recorded in \code{Modify} and holds 91 to 92\% throughout, so the limit is a property of the solve, not of
the molecular dynamics around it. ``Mesh'' is the reciprocal-space force evaluation; ``other'' pools the pair, bond, neighbor, communication and output time, each under 1.1\%.}
\label{fig:gpuscale}
\end{figure*}

These results are for one node. Whether the same solve carries across nodes is a separate question, and the answer sets the range of use of the present implementation. On 
Perlmutter, whose GPU nodes carry four NVIDIA A100 devices, we repeated the benchmark at 331,776 and 786,432 atoms on one, two and four nodes, with the device solve and 
GPU-aware MPI throughout (Figure~\ref{fig:gpuscale}a). A single A100 node is 2.1 and 2.6 times a V100 node on the same deck, at 3.41 and 1.47 timesteps per second. Beyond 
it the returns are small: two nodes are marginally \emph{slower} than one at both sizes, 3.38 against 3.41 and 1.39 against 1.47, and four nodes reach only
4.47 and 1.82, which is 1.3 and 1.2 times the single-node rate at a parallel efficiency of about a third.

The timing breakdown localizes the limitation. The charge solve is recorded in \code{Modify}, and it takes 91 to 92\% of the timestep at every size 
and every node count (Figure~\ref{fig:gpuscale}b), while the communication LAMMPS reports falls from 0.58 to 0.19\% as the atom count per device drops. 
The iteration count is also unchanged, at 15 conjugate-gradient iterations per step in all six runs, so the solve is not doing more work on more nodes. What grows is 
the cost inside it: each iteration carries a global reduction for its inner products and a distributed transform for the reciprocal term, roughly thirty collectives 
per timestep, and those are latency-bound once they cross the interconnect. The standard task breakdown cannot resolve this, because the solve's own communication is 
charged to the fix rather than to \code{Comm}.

We therefore recommend the current implementation for single-node device runs, where it is about twice a CPU node, and not as a route to strong scaling across nodes. 
Reducing the number of collectives per step, by a communication-avoiding Krylov variant or, for thermostatted runs, by accepting the ASPC's looser per-step criterion, is the natural next step.

\begin{table}[t]
\caption{Throughput of the water benchmark with the full charge solve at every step, in timesteps
per second, on one Expanse CPU node (128 cores) and on one and four V100 GPUs with the device solve and
GPU-aware MPI.}
\label{tab:gpu}
\begin{tabular}{rccc}
\toprule
Atoms & CPU node & 1 V100 & 4 V100 \\
\midrule
12,288  & 21.14 & 17.27 & 23.37 \\
98,304  & 3.172 & 1.458 & 6.269 \\
331,776 & 0.849 & 0.275 & 1.643 \\
786,432 & 0.305 & 0.107 & 0.568 \\
\bottomrule
\end{tabular}
\end{table}

Table~\ref{tab:propagators} gives the cost of four charge-dynamics modes on one system in the microcanonical (NVE) ensemble. The system is the 64 molecules of the water benchmark 
at 1.00~g\,cm$^{-3}$, rigid, at a 1~fs step, carried through 200~ps of NVE dynamics after 100~ps of thermostatted equilibration at 298~K. The shielding correction of the pair style 
extends to the solve cutoff of 10~\AA{}, and the Lennard-Jones term is shifted to zero there. The cutoff exceeds half the 12.4~\AA{} box; LAMMPS includes every periodic image within it, so the operator is exact for the periodic system, and the small box serves only to compare the modes on one system. Drifts are the slope of the total energy per rigid-body degree of freedom, expressed as 
a temperature (K\,ns$^{-1}$), with the standard error from ten contiguous segments.

\begin{table}[t]
\caption{The four charge-dynamics modes on one system, 64 molecules of the water benchmark over 200~ps of
microcanonical dynamics, on the deck described in the text, at a relative residual of $10^{-7}$. The time
per step is on one core, the mean of three repeats for the two solves ($\pm1$~ms). For the two modes that
advance the charges a drift does not describe the trajectory, and the last column states what happens
instead.}
\label{tab:propagators}
\begin{tabular}{lccl}
\toprule
Charge dynamics & Matrix-vector & Time per & Microcanonical behavior \\
                & products/step & step (ms) & over 200~ps \\
\midrule
Full solve                & 17.94 & 10.3 & $+20 \pm 21$~K\,ns$^{-1}$ \\
Solve seeded from history & 13.52 & 9.4  & $+20 \pm 9$~K\,ns$^{-1}$ \\
Predictor--corrector      & 2.00  & 5.7  & $-1145 \pm 15$~K\,ns$^{-1}$; $\mu$ 2.76 $\to$ 2.96~D \\
Extended Lagrangian       & ---   & 5.5  & $T$ falls from 270 to 85~K in 10~ps; $\mu \to 3.52$~D \\
\bottomrule
\end{tabular}
\end{table}

The full solve conserves energy to within the resolution of a 200~ps run, and is insensitive to tightening the tolerance: the drift is $+6 \pm 15$, $+20 \pm 21$ and 
$+1 \pm 13$~K\,ns$^{-1}$ at relative residuals of $10^{-5}$, $10^{-7}$ and $10^{-9}$, and $-7 \pm 18$~K\,ns$^{-1}$ at $10^{-9}$ and a 0.5~fs step.
The total energy fluctuates by 12 to 21~meV about its trend in all four runs. What does matter is the cutoff. The same system 
with the shielding correction truncated at 5~\AA{} against the solve's 10~\AA{}, a deck the initialization check accepts because the residual correction there is small, 
drifts by $+54 \pm 32$~K\,ns$^{-1}$ and its total energy fluctuates by 64~meV, three to five times more. We therefore recommend the full solve, or the solve seeded from history, with the shielding 
correction carried to the solve cutoff.

Computational cost tracks the iteration count, as expected. Seeding the solve from the previously converged charges saves a quarter of the matrix-vector products, which on this 
small system is within the timing noise. The ASPC integrator needs 9 times fewer products than the full solve and runs 1.8 times faster, because the pair and mesh work it does 
not remove dominates at 2 products per step. Shared-memory threading of the solve adds less: on one MPI rank of a 5184-atom water box, four threads reduce the wall time by a factor of 1.23 and eight by 1.26, where two MPI ranks give 1.80, so ranks remain the first choice. Conversely, in the near-singular metallic limit, where 
the operator is poorly conditioned, a block incomplete-LU preconditioner holds the saddle-point solve of a 26-atom metal slab at three iterations as the hardness approaches zero, 
where a diagonal preconditioner needs 59 iterations at a hardness of 0.5~eV and fails to converge below 0.1~eV. At the default drop tolerance of $10^{-3}$ the count grows with system size; at $10^{-5}$ it stays at six 
or fewer from 26 to 1682 atoms.

We find a distinct failure mode for the two charge integrators that advance rather than solve the charges at every step, in NVE dynamics of this model at the documented settings (one charge mass and damping for the extended Lagrangian, the default acceptance tolerance for ASPC); we did not scan those settings. 
The ASPC cools the system while its mean molecular dipole rises. Its acceptance criterion, a relative residual below $10^{-2}$, admits every step: the accepted residuals sit near 
$5\times10^{-4}$, far above the solve tolerance, so the fallback to the full solve never triggers. The extended Lagrangian fails sooner and more severely, although the thermostatted 
equilibration does not distinguish it from the other modes. In both, energy leaves the nuclear motion and appears in the polarization, which acts as a sink for it. 
The lesson here is that satisfying the residual criterion and conserving energy are separate properties, so a trajectory propagated by either mode must be validated against a fully solved 
one, not by its own residual in NVE. A thermostat masks this failure, so the fidelity of a propagated trajectory can be checked only in the microcanonical ensemble.

% =====================================================================

\section{Comparison with related methods}
\label{sec:comparison}
% =====================================================================

Table~\ref{tab:lammpsqeq} places \samQEq{} among the charge-equilibration routes of LAMMPS, and it implements several features which are otherwise spread amongst various LAMMPS fixes 
and external codes. The functional of Eq.~\eqref{eq:functional} is that of QEq and EEM,\cite{Mortier1986,Rappe1991} and in its globally constrained mode \samQEq{}  reproduces their behavior, 
including the unlimited inter-fragment transfer that follows from equalizing one chemical potential over a whole system. It also reproduces the charge step of ReaxFF to 
about $10^{-9}\,e$ (Section~\ref{sec:reaxff}). The per-fragment-neutral mode reproduces the constrained fluctuating-charge model of Rick and co-workers,\cite{Rick1994} 
which has run with per-molecule constraints under particle-mesh electrostatics in biomolecular force fields for two decades,\cite{Patel2004} although as currently implemented 
the \samQEq{} projector acts inside a Krylov solve with the reciprocal term in the matrix-vector product rather than on extended-Lagrangian charge dynamics. The conductor construction 
in \samQEq{} is similar to the ELECTRODE model already in LAMMPS: the finite-width self-energy in the global mode is equivalent to the Thomas--Fermi approach of Scalfi and co-workers,\cite{Scalfi2020} 
whose Gaussian electrode charges carry that self-energy implicitly, and the ELECTRODE package solves it on the same particle mesh at constant potential or at constant charge.\cite{AhrensIwers2022} 
What \samQEq{} adds is a fluctuating-\emph{charge} electrolyte, constrained molecule by molecule from the connectivity, in the same charge-equilibration solve as a Gaussian-regularized conductor, within LAMMPS and with a selectable kernel. The ELECTRODE package keeps the electrolyte at fixed charges; MetalWalls polarizes it through induced dipoles rather than fluctuating charges;\cite{Coretti2022} and the fragment-constrained model of Peeketi \emph{et al.} learns its fragments and electronegativities inside a machine-learned potential for an oxide electrode, without a conductor self-energy.\cite{Peeketi2026} 
A constant-potential/constant-current extension to \samQEq{} is another natural next step.
 
\begin{table}[h]

\caption{Charge-equilibration routes in LAMMPS. ``Long range'' is how interactions beyond the cutoff enter the
charge solve; ``conductor'' is whether the self-energy of a Gaussian charge can stabilize a hardness near zero.}
\label{tab:lammpsqeq}
\small
\begin{tabular}{@{}p{0.25\textwidth}p{0.17\textwidth}p{0.15\textwidth}p{0.20\textwidth}p{0.13\textwidth}@{}}
\toprule
Route & Kernel & Long range & Constraint & Conductor \\
\midrule
\code{qeq/point}, \code{qeq/shielded}, \code{qeq/dynamic}, \code{qeq/fire} & point or shielded & cutoff & global & no \\
\code{qeq/slater} & Slater & Wolf or DSF & global & no \\
\code{qeq/reaxff}, \code{acks2/reaxff}, \code{qtpie/reaxff} & shielded, tapered & cutoff & global (ACKS2, QTPIE attenuate transfer) & no \\
\code{electrode/conp}, \code{electrode/conq} & Gaussian & particle mesh & per electrode; electrolyte fixed & yes \\
\code{qeq/sam} (this work) & cube-root, Gaussian, Slater & particle mesh, inside the solve & per molecule and per conductor, or global & yes \\
\bottomrule
\end{tabular}
\end{table}

\samQEq{} builds on the QTPIE\cite{Chen2007} and split-charge\cite{Nistor2006} QEq formulations, and differs in the variables retained. Those methods transform to charge-transfer 
or bond variables, which delivers the correct dissociation asymptotics directly; \samQEq{} remains in atomic charges and applies the restriction through the constraint. 
The atomic-charge formulation keeps the solve in the same variables the pair styles and the reciprocal-space solver already use, which is what makes the
evaluation of the reciprocal term inside the matrix-vector product and the composition of Section~\ref{sec:coupling} possible. The cost is that polarization is represented 
on atoms rather than along bonds, a limitation shared by all atom-centered charge models.\cite{Gergs2021}

Compared to ACKS2,\cite{Verstraelen2013} which derives its response kernel from Kohn--Sham theory and thereby obtains the correct macroscopic polarizability
scaling, \samQEq{} is empirical: the separation between intramolecular and intermolecular response is imposed rather than derived. The two are nonetheless complementary: 
the ACKS2 bond-softness kernel is a run-time setting in \samQEq{} and restores linear scaling (Section~\ref{sec:limitations}), so a system whose longitudinal response 
matters should use it. Obtaining both the correct scaling and the correct magnitude requires a reparameterization outside the scope of this paper.
Finally, \samQEq{} provides the same ingredients as recent machine-learned QEq approaches,\cite{Ko2021,Cheng2025,Peeketi2026} but does so as an abstraction layer independent 
of the short-range potential itself. Section~\ref{sec:mlip} provides a perspective on how its Gaussian kernel maps onto the Latent Ewald Summation approach.\cite{Cheng2025,Kim2025LES}

% =====================================================================
\section{Conclusions}
% =====================================================================

In summary, we have developed \samQEq{}, a charge-equilibration framework that holds molecules and conductors in one energy functional and one projected solve. One neutrality 
constraint per molecule gives the insulating limit of fluctuating-charge models, while a finite-width self-energy of a Gaussian charge, narrower than the conductor's own, keeps the
conductor's part of the operator positive definite down to the conductor limit of the model. On a gold slab carrying a water film, of the four combinations of constraint and self-energy only the two together keep the operator definite down to the conductor limit without charge transfer between molecules, 
and the metal screens a point charge by a constant amount beyond 4~\AA{}. The shielding kernel and the long-range treatment are run-time settings 
under full particle-mesh electrostatics inside the matrix-vector product. We validated the implementation by reproducing the published Slater-overlap TIP4P-FQ water model and the ReaxFF 
charge step with no refit. The Kokkos device implementation shows that a four-GPU node runs at about twice the throughput of a CPU node, and the package composes with fixed-charge and reactive 
force fields. We show that a converged, neutral charge set does not ensure the intended solve, and we document the initialization checks that catch the operator substitutions a converged
residual cannot reveal.

More generally, the molecular and conductor limits of charge equilibration are two conditions on one operator, and both can be imposed in the same solve.
For the Gaussian kernel, Eq.~\eqref{eq:gram} bounds the conductor's contribution to that operator from below at any hardness, up to the mesh residual, so the conductor limit is well posed by
construction rather than by tuning. An interface between an electrolyte and an electrode contains both limits, and a model that is well posed in only one of them
cannot describe it.

This framework defines the next set of problems. The first is the physics of the framework itself: native fluctuating-charge water, the constraint axis between
the molecular and metallic limits, conductors as a function of field strength and slab thickness, and the longitudinal polarizability of chain molecules. The frontier
levels that every parameter file already carries are the basis for charge transfer between fragments gated by those levels, so that transfer switches on only where it
is chemically allowed. Constant-potential electrodes, redox and grand-canonical charge exchange at electrochemical interfaces build on the same solve. The initialization
checks are the starting point for a systematic treatment of silent operator substitutions in charge solvers. Coupling to machine-learned potentials that predict the
electronegativities requires no derivative of the charges (Section~\ref{sec:mlip}), and Drude-like shells enter the solve as fixed charges through their field in
$\mathbf{b}$ (Section~\ref{sec:constraint}). \samQEq{} thus makes molecules, conductors and the interfaces between them one electrostatics problem, available to any
LAMMPS potential through a single constrained solve.

% =====================================================================
\section*{CRediT authorship contribution statement}
\textbf{Nicholas Solan:} Data curation, Formal analysis, Investigation, Methodology, Validation, Writing -- review \& editing.
\textbf{David Sun:} Data curation, Formal analysis, Investigation, Methodology, Validation, Writing -- review \& editing.
\textbf{Tod A. Pascal:} Conceptualization, Formal analysis, Funding acquisition, Methodology, Project administration, Software, Supervision, Writing -- original draft, 
Writing -- review \& editing.

% =====================================================================
\section*{Declaration of Competing Interest}
The authors declare that they have no known competing financial interests or personal relationships that could have appeared to influence the work reported in this paper.

% =====================================================================
\section*{Data Availability}
The \samQEq{} source, the regression suite, the patches and the example decks, including the runnable example \code{kernel\_comparison}, which exercises both kernels of Table~\ref{tab:kernelcmp} in a 40~\AA{} cell, 
are available at \url{https://github.com/atlas-nano/codes/tree/main/samqeq}; the version described here, 1.0.2, is archived at Zenodo.\cite{Pascal2026samqeq} The data and scripts that reproduce the figures are available at \url{https://github.com/atlas-nano/manuscript-data/tree/main/samqeq-molecules-conductors}.

% =====================================================================
\section*{Acknowledgments}
This work was supported by the U.S.\ National Science Foundation through grant CBET-2145144. This research used resources of the National Energy Research
Scientific Computing Center, a DOE Office of Science User Facility supported by the Office of Science of the U.S. Department of Energy under Contract
No.\ DE-AC02-05CH11231 using NERSC award DDR-ERCAP0039121. This work used Expanse at the San Diego Supercomputer Center through allocation
TG-DMR190106 from the Advanced Cyberinfrastructure Coordination Ecosystem: Services \& Support (ACCESS) program, which is supported by U.S.\ National Science
Foundation grants \#2138259, \#2138286, \#2138307, \#2137603 and \#2138296. T.A.P.\ acknowledges support from the Alfred P.\ Sloan Foundation.

\section*{Supplementary Material}
The Supplementary Material provides the per-fragment projector; the long-range reciprocal-in-matrix-vector construction; the solvers and their
convergence criteria; the dense diagonalization of the constrained operator and its check against the solved charges; the water model parameters and
protocols; the construction and protocol of the molecule--conductor interface; the regression suite; the device benchmark; and the sensitivity to the charge-solve tolerance.

\bibliographystyle{elsarticle-num-names}
\bibliography{refs}

\end{document}

\let\WriteBookmarks\relax

\shorttitle{Supplementary Material: samQEq for molecules and conductors}
\shortauthors{Solan, Sun and Pascal}
\title[mode=title]{Supplementary Material for samQEq: universal charge equilibration for molecules and conductors}

\author[1]{Nicholas Solan}
\fnmark[1]

\author[1]{David Sun}
\fnmark[1]

\author[1,2]{Tod A. Pascal}[orcid=0000-0003-2096-1143]
\cormark[1]
\ead{tpascal@ucsd.edu}
\affiliation[1]{organization={ATLAS Materials Physics Laboratory, Aiiso Yufeng Li
            Family Department of Chemical and Nano Engineering,
            University of California San Diego},
            city={La Jolla},
            citysep={,},
            postcode={92093},
            state={CA},
            country={USA}}
\affiliation[2]{organization={Program in Materials Science and Engineering,
            University of California San Diego},
            city={La Jolla},
            citysep={,},
            postcode={92093},
            state={CA},
            country={USA}}
\cortext[1]{Corresponding author}
\fntext[1]{These authors contributed equally to this work.}

\maketitle

\tableofcontents
\newpage

% =====================================================================
\section{Computational approach}
% =====================================================================

\subsection{The per-fragment projector}
\label{sec:projector}

Fragments were identified by molecule ID and the projector of Eq.~\eqref{M-eq:projector} was applied to the matrix-vector product output, to the search directions and to 
the right-hand side, so that the Krylov method sees the symmetric operator of Eq.~\eqref{M-eq:solve}. In the absence of molecule IDs, the code treats every atom in 
the group as a single fragment. We formed the per-fragment sums over the occupied fragments rather than over the largest molecule ID, which keeps the cost independent of 
how molecule IDs are numbered; the sums, and hence the charges, are the same either way.

\subsection{Long-range electrostatics inside the matrix-vector product}
\label{sec:longrange}

Both long-range modes of Section~\ref{M-sec:longrange} of the main text were run under a thermostat (Table~\ref{tab:si-protocols}). The intramolecular-only mode sits
closer to the polarization catastrophe (Section~\ref{M-sec:longrange} of the main text), at which the charges polarize without bound.\cite{Rick1994}
We measured the grid-dependent reciprocal self-term once per set of mesh dimensions and Ewald parameter, rather than at every run start, 
which avoids recalibrating on a grid that has not yet been finalized. Each of the sixteen default probes pairs one of the lowest global tags with its nearest partner 
between 0.1~\AA{} and 2.5 mesh spacings. Every recalibration is checked against the first at the same mesh, and the run aborts at a 2\% departure. The exact per-atom 
route subtracts the analytic periodic-image term from each atom's own grid self-potential and is re-evaluated at every solve.

% =====================================================================
\section{The solvers}
% =====================================================================

\subsection{Projected MINRES and the stall gate}
\label{sec:minres}

MINRES\cite{Paige1975} is provided for indefinite operators. It does not replace the group-membership check of Section~\ref{M-sec:checks} of the main text, which refuses 
such a solve at initialization. We solved $\Amat \mathbf{x} = \mathbf{b}$ by the MINRES algorithm unpreconditioned, where $\mathbf{x}$ is the vector of charge 
deviations and $\mathbf{b}$ the projected right-hand side. The approach builds a Lanczos\cite{Lanczos1950} tridiagonalization of $\Amat$ and applies a Givens QR factorization to it, 
and minimizes $\|\mathbf{b} - \Amat\mathbf{x}\|$ over the Krylov subspace. The only requirement is symmetry; positive definiteness is not assumed. The operator is singular by
construction, since the projector annihilates one direction per fragment, and MINRES carries no general guarantee on a singular system. Indeed, iterations of an inconsistent solve 
can grow without bound while the residual norm settles at its least-squares floor, and a minimum-norm solution needs MINRES-QLP\cite{Choi2011} rather than MINRES. Neither 
applies here. The right-hand side is projected, the operator maps the constrained subspace into itself, and the iteration starts inside it, so the system MINRES actually
sees is consistent, and the null space is never excited.

We initialized $\beta_1 = \|\mathbf{b}\|$, the unnormalized Lanczos vectors $\mathbf{r}_1 = \mathbf{r}_2 = \mathbf{b}$, the direction vectors $\mathbf{w} = \mathbf{w}_2 = 0$, 
$\bar\phi = \beta_1$, $c = -1$ and $s = 0$. Each iteration forms the Lanczos vector $\mathbf{v} = \mathbf{r}_2/\beta$, applies the operator, orthogonalizes
against the two previous Lanczos vectors to give the tridiagonal entries $\alpha$ and $\beta$, and then updates the QR factors as
\begin{align}
\delta &= c\,\bar{d} + s\,\alpha, &
\bar{g} &= s\,\bar{d} - c\,\alpha, \label{eq:si-givens1}\\
\epsilon &= s\,\beta, &
\bar{d} &= -c\,\beta, \label{eq:si-givens2}\\
g &= \sqrt{\bar{g}^2 + \beta^2}, &
c &= \bar{g}/g, \quad s = \beta/g, \label{eq:si-givens3}\\
\phi &= c\,\bar\phi, &
\bar\phi &\leftarrow s\,\bar\phi , \label{eq:si-givens4}
\end{align}
where $c$ and $s$ are the cosine and sine of the current Givens rotation, $\bar{d}$ and $\bar{g}$ are the running diagonal and subdiagonal of the factorized tridiagonal, 
$\delta$ and $\epsilon$ are the two off-diagonal carries, $g$ is the rotation norm, $\phi$ is the current solution coefficient, and $\bar\phi$ is the residual norm. We 
then applied the three-term recurrence for the direction vectors and accumulated $\mathbf{x} \leftarrow \mathbf{x} + \phi\,\mathbf{w}$.

Because $\bar\phi$ is the residual norm, the stopping test $\bar\phi/\|\mathbf{b}\| \le \mathrm{tol}$ is a true relative residual and is dimensionally invariant, unlike 
a preconditioned conjugate-gradient criterion built from $\mathbf{r}^{\!\top}\mathbf{D}^{-1}\mathbf{r}$ (with $\mathbf{D}$ the diagonal preconditioner), which carries the 
units of the preconditioner. Because the matrix-vector product projects its output, every Lanczos vector remains in the constrained subspace, so the method solves the 
projected problem with no further machinery.

\paragraph{The stall gate.}
From Eq.~\eqref{eq:si-givens4}, $\bar\phi \leftarrow s\,\bar\phi$, so $s$ measures the per-iteration residual reduction directly. A value of $s$ close to unity means only 
that the residual has stopped falling (consistent with growth along a near-null direction) but does not strictly establish it. Stagnation can follow from clustered eigenvalues, 
from loss of Lanczos orthogonality in finite precision, and, were the projected right-hand side ever inconsistent, from the residual settling at its least-squares floor. The
criterion is therefore an empirical detector for growth along a near-null direction of a near-singular operator in finite
precision, rather than a diagnosis of its cause. The gate is a heuristic safeguard, a bounded early termination rather than a regularized solve. We applied it on $s > 0.995$ (i.e., on less than half a percent reduction per iteration), and on a 
charge magnitude above a cap. When both held, we terminated the iteration and returned the bounded, regularized iterate. Gating on the charge magnitude alone proved to be incorrect: 
MINRES here is non-monotonic, so a healthy converging solve may transiently exceed the cap, and truncating it would return a non-finite result for a system that otherwise solves correctly.

\paragraph{Deployment.}
We left projected conjugate gradient as the default path. When a solve returns a maximum charge magnitude above a threshold, we re-solve that same step with MINRES with 
the truncation cap engaged. The user may also select MINRES directly (Section~\ref{M-sec:controls} of the main text) if desired.

\subsection{The eigenvalue estimate and the configuration-smooth ridge}
\label{sec:lanczos}

We estimated the smallest eigenvalue of the projected operator from an $m$-step Lanczos tridiagonalization and obtained the smallest eigenvalue of the resulting tridiagonal 
$\mathbf{T}$ by Sturm-sequence bisection. The Sturm count is
\begin{equation}
\nu(\sigma) = \#\left\{ \text{negative pivots of }
\mathbf{L}\mathbf{D}\mathbf{L}^{\!\top}(\mathbf{T} - \sigma\mathbf{I}) \right\}
= \#\left\{ \lambda_i(\mathbf{T}) < \sigma \right\},
\label{eq:si-sturm}
\end{equation}
where $\sigma$ is the trial shift, $\mathbf{I}$ the identity, $\mathbf{L}\mathbf{D}\mathbf{L}^{\!\top}$ the triangular factorization with unit-diagonal $\mathbf{L}$ and pivots 
on the diagonal of $\mathbf{D}$, and $\lambda_i$ the eigenvalues of $\mathbf{T}$. In short, $\lmin$ is the smallest $\sigma$ for which $\nu(\sigma) \ge 1$, which we located by
bisection from Gershgorin spectral bounds.\cite{Parlett1998}

The bisection returns the smallest eigenvalue of $\mathbf{T}$ to the requested tolerance, but that is a Ritz value, a Rayleigh quotient on the Krylov space, and therefore an \emph{upper} bound 
on $\lmin$ of the full operator. The error is one-sided in the direction that matters: an unconverged $m$ can only miss a negative mode, never invent one. On one interface
cell, $m = 15$ returns $+2.17$~eV where $m \ge 60$ gives $-0.31$~eV. The implementation therefore also reports the $(m-10)$-step Ritz value from the same tridiagonal, 
which costs nothing, and warns once if the two estimates still differ.

We then set a Tikhonov ridge $\rho$ such that $\lmin(\Amat + \rho\mathbf{I})$ meets a specified floor.\cite{Tikhonov1977,Hansen1998} Because $\rho$ is a continuous function of 
the configuration when it is recomputed at every step, the charges and hence the forces then vary continuously through a near-critical configuration. Recomputed every $N$ steps, $\rho$ is piecewise constant and the forces jump at each update. Because the Ritz value is an upper bound, an unconverged estimate can also set $\rho$ too small to reach the floor. A discrete escalation, raising the ridge in steps when a solve misbehaves, 
instead makes the forces discontinuous. The ridge is not part of the reported energy, so forces are not conservative while it is active.

The Lanczos estimate costs $m$ matrix-vector products. Recomputing it every $N$ steps rather than every step reduced the wall time of a 512-molecule water run under the full solve in the microcanonical ensemble by a factor of 1.5 at $N = 10$, with the same energy drift.

% =====================================================================
% =====================================================================
\section{Dense diagonalization of the benchmark operator}
\label{sec:dense-native}

We took the final configuration of a 30~ps isothermal--isobaric run of the shipped deck of the benchmark water model
(512 molecules, 298~K, 1~bar, pinned self-term), solved its charges once to a relative residual of
$10^{-10}$ with the conjugate-gradient path and the fallback disabled, and rebuilt the constrained operator
from the geometry and the parameter file. Here, $H_{ii} = \eta_i$, with the cube-root kernel at
$\gamma = 1.11$~\AA$^{-1}$ for every pair inside the 10~\AA{} solve cutoff, and the reciprocal-space term as
an Ewald sum at the splitting parameter and self-term the solve applies, 0.33957~\AA$^{-1}$ and 0.37933~\AA$^{-1}$ (in units of $K_e$).
The rebuilt operator reproduces the solved charges to $3.3\times10^{-6}\,e$ at a particle-mesh accuracy of
$10^{-6}$ and to $2.7\times10^{-8}\,e$ at $10^{-8}$. On the 1024-dimensional constrained subspace it has
$\lmin = 1.358$~eV and $\lambda_{\max} = 10.906$~eV, a condition number of 8.03 and no negative
eigenvalue. The 512 exact zero modes of the projector, one per molecule, are removed by the change of
basis: 1536 atomic degrees of freedom carry 512 neutrality constraints, leaving
the 1024 reported here. The real-space kernel alone, truncated at 10~\AA{}, has 273
negative eigenvalues, so the measurement keeps the reciprocal term. A fresh single point on the same
configuration retunes the splitting parameter to 0.34576~\AA$^{-1}$; used with the pinned self-term, the retuned splitting
shifts every eigenvalue by $+0.10$~eV and lowers the solved dipole to 2.62~D.

% =====================================================================
\section{Water models: parameters and protocols}
\label{sec:water}
% =====================================================================

\subsection{Zero-refit port of a published model}

We used the exact Slater kernel of Table~\ref{M-tab:kernels} with the published exponents (Table~\ref{tab:si-params}) and no parameter adjustment, in the
intramolecular-shielding-only mode that matches that model's structure,\cite{Rick1994} and repeated the liquid simulation in the all-pairs-shielded mode over three independent trajectories
(Table~\ref{tab:si-protocols}). Section~\ref{M-sec:port} of the main text reports the dipoles.

\subsection{Parameter sets}
\label{sec:si-params}

The parameter-file format is given in Section~\ref{M-sec:controls} of the main text. The fourth column
is read according to the active kernel, so Table~\ref{tab:si-params} labels it per model.
The Slater exponents of the ported model are the published values converted from
atomic units, $\zeta_{\rm O} = 1.63/0.529177 = 3.0803$~\AA$^{-1}$ for the
oxygen and M-site $2s$ form factor and $\zeta_{\rm H} =
0.90/0.529177 = 1.7007$~\AA$^{-1}$ for the hydrogen $1s$. Both O and M carry the
oxygen form factor, following the convention of the original model, in which the
M site borrows the oxygen exponent although it sits 0.15~\AA{} off the oxygen
nucleus. Nothing else was adjusted.

\begin{table}[h]
\caption{\samQEq{} per-type parameters for every model reported
in this work. $\chi$ is in eV and $\eta$ in eV\,$e^{-2}$; the fourth column is
read according to the active kernel, as named per model. $q^0$ is the reference
charge about which the on-site expansion is taken. The frontier levels
$\varepsilon_{\rm HOMO} = -7$~eV and $\varepsilon_{\rm LUMO} = 3$~eV are carried by every type and enter no calculation reported here.}
\label{tab:si-params}
\begin{tabular}{llrrrr}
\toprule
Model & Type & $\chi$ & $\eta$ & 4th col. & $q^0$ \\
\midrule
\multicolumn{6}{l}{\emph{Ported TIP4P-FQ (bare intermolecular), Ref.~\citep{Rick1994}, kernel \code{slater}, $\zeta$ (\AA$^{-1}$)}} \\
& O & 0.00   & 100.00 & 3.0803 & 0.0 \\
& H & 4.50   & 15.31  & 1.7007 & 0.0 \\
& M & 7.47   & 16.11  & 3.0803 & 0.0 \\
\midrule
\multicolumn{6}{l}{\emph{Benchmark water (native fluctuating-charge model, fitted to the gas dipole and liquid density), kernel \code{cbrt}, $\gamma$ (\AA$^{-1}$)}} \\
& O & 6.7721 & 15.1145 & 1.11 & 0.0 \\
& H & 4.5000 & 16.1595 & 1.11 & 0.0 \\
\bottomrule
\end{tabular}
\end{table}

\subsection{Simulation protocols}
\label{sec:si-protocols}

Table~\ref{tab:si-protocols} collects the box, ensemble and run length of each
production simulation reported in the main text. Both models use a
rigid-molecule integrator for the nuclear degrees of freedom, and all production charge solves use a convergence tolerance of $10^{-5}$ on the relative residual, which
Section~\ref{sec:si-tolscan} shows lies well inside the property plateau. Reciprocal-space sums use
the particle-mesh solver at a relative accuracy of $10^{-5}$ for the ported model
and $10^{-6}$ for the native model.

\begin{table}[h]
\caption{Production protocols. $N$ is the number of water molecules, $L$ the
cubic box edge. Thermostat and barostat damping times follow the LAMMPS
convention, in picoseconds. The ported model is reported against
Ref.~\citep{Rick1994} rather than against experiment. The shipped-deck dipole of Section~\ref{M-sec:port} of the main text is
the mean over the eight runs of the first row, with the standard error over runs; the all-pairs-shielded dipole is the mean
over the three runs of the second row, with the standard error over runs.}
\label{tab:si-protocols}
\small
\setlength{\tabcolsep}{4pt}
\begin{tabular}{@{}llc>{\raggedright\arraybackslash}p{0.27\textwidth}>{\raggedright\arraybackslash}p{0.21\textwidth}@{}}
\toprule
Model & $N$, $L$ (\AA) & $\Delta t$ (fs) & Ensemble, damping & Run \\
\midrule
Ported, shipped deck & 256, 19.713 & 0.25 & NVT 298~K, $\tau_T$ 0.02 or 0.1 & eight seeds $\times$ (20~ps equil.\ $+$ 100~ps) \\
Ported, all-pairs-shielded & 256, 19.713 & 0.25 & NVT 298~K, $\tau_T$ 0.02 & three seeds $\times$ 10~ps \\
Native & 512, 24.800 & 1.0 & NPT 298~K, 1~bar; $\tau_T$ 0.1, $\tau_P$ 1.0 &
  15~ps equil.\ $+$ 15~ps \\
\bottomrule
\end{tabular}
\end{table}

The ported liquid was run from the shipped deck with eight velocity seeds, each equilibrated for 20~ps at
$\tau_T = 0.02$~ps (the shipped value) and then continued for 100~ps of production at $\tau_T = 0.02$ and at 0.1~ps,
with the grid self-term pinned at its calibrated value. Seven of the eight starts give a mean molecular dipole of
2.6299~D at 0.02~ps and 2.6298~D at 0.1~ps, so the strong coupling does not bias it, and $2.630 \pm 0.004$~D over the
seven (standard deviation over starts; standard error 0.0014~D). The eighth settles during its equilibration into a
network with fewer hydrogen bonds, 3.53 per molecule against 3.59 to 3.65 in the others (intermolecular O$\cdots$H
within 2.4~\AA{}), and holds a dipole of 2.55~D through 100~ps at both couplings. A Born--Oppenheimer re-solve of its
configurations reproduces the propagated dipole to $10^{-4}$~D, so the lower value belongs to the structure, not to
the charge propagation.

% =====================================================================
% =====================================================================
\section{Molecules and a conductor in one solve: building and simulation protocol}
\label{sec:twolimits}

We built the interface of Section~\ref{M-sec:twolimits} of the main text from two systems distributed with the package, with no
parameters fitted to explicitly describe the interfacial interactions. 
The metal is the 240-atom Au(111) slab of Section~\ref{M-sec:conductor}: five layers of 48
atoms in a $17.651 \times 20.381 \times 60$~\AA{} cell, layers at $z = 10.0$ to 19.608~\AA{}, 
periodic in the plane and free normal to it, with a nearest-neighbor distance of 2.942~\AA{} (a lattice constant of 4.16~\AA{}, 2\% above the experimental 4.078~\AA{}; the Thomas--Fermi conversion of the main text uses the experimental atomic density), $\chi = 5.31$~eV, a Gaussian radius 
of 1.618~\AA{} and a
self-energy width of 0.5~\AA{} when the self-energy is on. The water is the Gaussian-kernel port of
4-site TIP4P-FQ distributed with the package (\code{examples/tip4pfq\_liquid}), whose oxygen is uncharged and whose
hydrogen and massless M sites carry the fluctuating charges. Its Gaussian radii, 0.404355~\AA{} for H
and 0.315328~\AA{} for M, were fitted so that the kernel reproduces the model's own intramolecular Coulomb
integrals at the rigid geometry, $J_{\rm HH}$ at 1.5139~\AA{} and $J_{\rm MH}$ at 0.8735~\AA{}. The
hardnesses and electronegativities are those of Rick \emph{et al.}\cite{Rick1994} Every site uses the Gaussian kernel with the default
$\lambda$; the pair cutoff is 12~\AA{}, the solve cutoff 10~\AA{}, and the particle mesh is that of the bare
slab (\code{pppm/samqeq}, accuracy $10^{-6}$, \code{kspace\_modify slab 3.0 mesh 24 28 80 gewald 0.35}).

We performed a single solve at a fixed configuration, on one rank, to a relative residual tolerance of
$10^{-8}$, with no applied field. Under the global constraint all atoms share one fragment; under
per-molecule constraints each water molecule is its own fragment and the slab another. We scanned the gold
hardness over 5.172, 2.5, 1.0, 0.3 and 0.05~eV; the parameter file requires $\eta > 0$, and 0.05~eV stands as
the conductor limit. For the spectrum we recovered the operator, column by column, by
applying its own matrix-vector product (diagonal, real-space kernel and reciprocal term with the grid
self-term removed) to each unit vector of the 483 solved sites. We projected it onto an orthonormal basis of
the constrained subspace (482 dimensions under the global constraint, 401 under per-molecule constraints),
and diagonalized it densely. The recovered operator is symmetric to $10^{-15}$ relative accuracy and, solved
directly, reproduces the charges of the code to within $2.1\times10^{-7}\,e$ at every point. Its smallest
eigenvalue agrees with the code's 200-step Lanczos estimate (Section~\ref{sec:lanczos}) to $10^{-3}$~eV. 
The molecular reference is the same film with the slab deleted. Molecular dipoles are taken about the oxygen. 
Under the global constraint the molecules carry net charge and their dipoles are origin-dependent.

For the screening test we replicated the slab three times in each lateral (xy) direction (2160 atoms,
$52.95 \times 61.14$~\AA{}) and placed a fixed $+1\,e$ point charge a height $z$ above the top plane and a
fixed $-1\,e$ the same height below the bottom plane. Both test charge locations are at the lateral center, 
and both are outside the solve
group, so that they enter only through their field in $\mathbf{b}$. The cell therefore 
stays neutral. The mesh was
$72 \times 84 \times 80$. Since site charges are a poor measure of the screening (the Gaussian
densities of adjacent layers overlap), we integrated the Gaussian density of every gold site, at the width
of the pair kernel between gold sites, over the half-space above the slab's mid-plane. The exact image
energy of a point charge $q$ at height $z$ above a conducting plane at $z_0$, with the lateral periodicity
of the cell of area $A$, is $E = (K_e q^2/2)(2\pi/A)\,[\,d - \sum_{\mathbf{G} \ne 0} e^{-Gd}/G\,]$ with
$d = 2(z - z_0)$, the sum running over the in-plane reciprocal vectors. We then fitted $z_0$ to the interaction
energies for $z \geq 4$~\AA{}. The water-electronegativity offset shifted $\chi$ of H and M together by
$-2$ to $+2$~eV on the fixed film. The self-energy width was scanned from 0.3 to 2.378~\AA{} on the
five-layer slab of Section~\ref{M-sec:conductor} in the uniform field of 0.5~V\,\AA$^{-1}$. To decompose the excess screening, the same field test was repeated on slabs of 3, 5, 7, 9, 13 and 17 layers and at gold Gaussian radii from 0.5 to 2.4~\AA{}, and the point-charge test on 13 layers; every laterally uniform response was also reproduced, to $10^{-3}$, by a one-dimensional model of the layer charges with Gaussian-sheet interactions and the net on-site hardness $\eta + s - J(0)$.

For MD, we kept the water rigid (\code{fix rigid/nvt/small}, 298~K, damping 0.1~ps, bodies keyed
on a per-atom label so that the global-constraint arm can place all atoms in one fragment without merging
the bodies), held the gold fixed, and added a generic Au--O Lennard-Jones term ($\sigma = 3.0$~\AA{},
$\epsilon = 0.05$~eV) to keep the film on the slab. This term is not a fitted gold--water potential; it is sufficient for
testing the charge solve. Each arm ran 25,000 steps of 0.2~fs, the timestep of the distributed Gaussian TIP4P-FQ
example (\code{examples/tip4pfq\_liquid}), solving to a relative residual of $10^{-8}$ at every step. The metal-free control is the same film
with the slab deleted. Under \code{pair\_style hybrid/overlay} the \code{coul/shield/intra} coefficients
must be given for every pair of solved types, as \code{pair\_coeff * *} or one line per pair (Section~\ref{M-sec:checks} of
the main text); the fix refuses a deck that omits one and names the missing pairs.

The two single-solve arms at the ends of the hardness scan ship with the package as the regression cases \code{conductor\_water\_samqeq} and \code{conductor\_water\_qeq}. The data and scripts that reproduce the figures are at \url{https://github.com/atlas-nano/manuscript-data/tree/main/samqeq-molecules-conductors}.

\section{Regression suite}
\label{sec:tests}
% =====================================================================

Table~\ref{tab:si-tests} lists the \ncases{} regression cases of Section~\ref{M-sec:regression} of the main
text, grouped by what they pin. All \ncases{} pass at the source commit from which the package is distributed.

\begin{table}[h]
\caption{The regression cases, grouped by what they pin. Each is a LAMMPS input deck with its committed
reference charges; the names are those of the distributed test directory.}
\label{tab:si-tests}
\small
\begin{tabular}{@{}p{0.30\textwidth}p{0.64\textwidth}@{}}
\toprule
Group & Cases \\
\midrule
Water models and long-range paths & \code{spcfq\_gas\_dsf}, \code{spcfq\_gas\_ewald}, \code{spcfq\_liq\_ewald},
  \code{tip4pfq\_gas} \\
Shielding kernels and fixed charges & \code{slater\_gas\_kernel}, \code{shield\_special00},
  \code{ionfield\_cl\_water} \\
Charge dynamics & \code{spcfq\_xl}, \code{spcfq\_xl\_staircase}, \code{aspc\_corrector} \\
Constraint and set-up & \code{charge\_solid\_global}, \code{second\_run\_molinv},
  \code{recalib\_on\_grid\_change} \\
Grid self-term & \code{recip\_self\_peratom} \\
Slabs and the conductor limit & \code{slab\_ewald3dc}, \code{slab\_ewald\_sawtooth},
  \code{metal\_slab\_gself}, \code{gself\_cluster\_energy} \\
Molecules and a conductor in one solve & \code{conductor\_water\_samqeq}, \code{conductor\_water\_qeq} \\
Augmented solve & \code{bondsoft\_gas}, \code{metal\_ilu\_group} \\
\bottomrule
\end{tabular}
\end{table}

% =====================================================================
\section{Device benchmark}
\label{sec:si-gpu}
% =====================================================================

We built the benchmark from the water model of Section~\ref{M-sec:longrange} of the main text, replicated to 12,288, 98,304,
331,776 and 786,432 atoms, with two changes that keep the whole step on the device: we constrained bonds
and angles with SHAKE rather than integrating rigid bodies, for which no device integrator exists, and ran
at fixed volume at the lattice density of 1.004~g\,cm$^{-3}$ with a Nos\'e--Hoover thermostat at a 1~fs
timestep. All intramolecular pairs stay in the pair styles at full weight, as in the native model. We
solved the charges to the shipped tolerance of $10^{-5}$ at every step, or advanced them with 
ASPC with two corrector iterations, and ran 1000, 200, 100 and 50 steps at the four sizes.
We ran the CPU benchmarks on one Expanse CPU node with 128 MPI ranks bound to cores, and the GPU benchmarks with one
MPI rank per V100, the device solve enabled (\code{fix\_modify <id> devsolve on}), the Kokkos package
with Newton's third law on and half neighbor lists, and GPU-aware MPI. The MPI library is CUDA-aware at
run time. Before timing, host and device forces at fixed charges agreed to $10^{-11}$~eV~\AA$^{-1}$ on
one and two GPUs, and host and device charges agreed to $10^{-11}\,e$ after ten steps.

% =====================================================================
% =====================================================================
\section{Sensitivity to the charge-solve tolerance}
\label{sec:si-tolscan}
% =====================================================================

\begin{figure}[t]
\centering
\includegraphics[width=\textwidth]{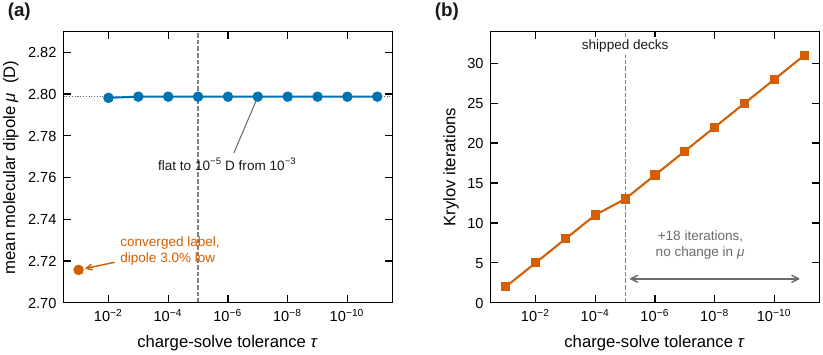}
\caption{Sensitivity of the solved dipole to the charge-solve tolerance
$\tau$, on the 512-molecule native fluctuating-charge water at fixed geometry, the final configuration
of a 30~ps isothermal--isobaric run solved with the self-term, Ewald splitting parameter and mesh of the
production run. $\mu$ is the mean molecular dipole of that configuration.
Tolerance loosens to the left. (a) The dipole is flat to $10^{-5}$~D from
$10^{-3}$ to $10^{-11}$ and within $5.4\times10^{-4}$~D at $10^{-2}$ (blue), and breaks at
$\tau = 10^{-1}$ (orange), where the solver reports convergence after two iterations and returns a value
3.0\% low. The dotted line marks the converged dipole, 2.798710~D. (b) Krylov
iterations against the same axis, growing by about three per decade. The dashed
line in both panels marks the $\tau = 10^{-5}$ of the shipped decks; the arrow
spans the range over which a further eighteen iterations change no reported
digit of $\mu$. Solved with the automatic fallback disabled.}
\label{fig:si-tolscan}
\end{figure}

We scanned the charge-solve tolerance over ten orders of magnitude on the water benchmark of
Section~\ref{M-sec:longrange} of the main text, on a configuration of 512 molecules taken from the end of a 30~ps
NPT run of the shipped deck, and recorded the mean molecular dipole and the number of
Krylov iterations at each setting. The solve was run with the automatic fallback disabled, so that a
struggling solve reports its own failure rather than being rescued, and with the self-term, Ewald splitting
parameter and mesh of the production run, so that the only quantity varying across the scan is the
tolerance. The splitting parameter must be held with the self-term (Section~\ref{sec:dense-native}). Charges were 
solved once at each setting on the fixed configuration, which isolates the accuracy
of the solve from any propagation of error through dynamics. The converged dipole is 2.7987~D.

The dipole was found to be insensitive to the tolerance across nine of the ten orders of magnitude
(Figure~\ref{fig:si-tolscan}), and the settings at $10^{-7}$ and tighter are identical to ten significant figures, 
while the cost keeps rising. The property plateau is therefore reached well before the
residual stops falling, and the shipped tolerance sits inside it.

The plateau does break, however, at the loosest setting tried, and the failure carries no diagnostic of its own:
the residual is below tolerance, the charges sum correctly and nothing in the output distinguishes the answer from
a converged one. A converged label is therefore necessary but not sufficient evidence of a correct solve.

% =====================================================================
\section{Code and data availability}
% =====================================================================

The \samQEq{} package, including the source, the regression suite, the patches and the example decks, is available at
\url{https://github.com/atlas-nano/codes/tree/main/samqeq}; version 1.0.2 is archived at
\url{https://doi.org/10.5281/zenodo.23629400}. The data and scripts that reproduce the figures are available at
\url{https://github.com/atlas-nano/manuscript-data/tree/main/samqeq-molecules-conductors}.

\bibliographystyle{elsarticle-num-names}
\bibliography{refs}